\documentclass[twocolumn,iop,revtex4,usenatbib]{openjournal}
\usepackage[T1]{fontenc}

\usepackage[english, english]{babel}

\DeclareRobustCommand{\VAN}[3]{#2}
\let\VANthebibliography\thebibliography
\def\thebibliography{\DeclareRobustCommand{\VAN}[3]{##3}\VANthebibliography}

\usepackage{algorithm} 
\usepackage[noend]{algpseudocode} 

\usepackage[bitstream-charter]{mathdesign}
\usepackage[usenames]{color}
\usepackage{xcolor}
\usepackage{soul}
\usepackage{multirow}
\usepackage{graphicx}
\usepackage{amsmath}
\usepackage{nicefrac}
\usepackage{microtype}
\usepackage{amssymb}
\usepackage{bbding}
\usepackage{csquotes} 
\usepackage{float}
\usepackage{xurl}
\usepackage{adjustbox}
\usepackage{comment}
\usepackage[hyperfootnotes=false]{hyperref}
\usepackage{tensor} 
\usepackage{orcidlink} 

\hypersetup{colorlinks=true, citecolor=blue, linkcolor=blue, urlcolor=blue}
\definecolor{hotpink}{RGB}{255, 105, 180}

\definecolor{orcidlogocol}{HTML}{A6CE39}

\DeclareSymbolFont{usualmathcal}{OMS}{cmsy}{m}{n}
\DeclareSymbolFontAlphabet{\mathcal}{usualmathcal}

\newcommand{\ltsima}{$\; \buildrel < \over \sim \;$}
\newcommand{\lsim}{\lower.5ex\hbox{\ltsima}}
\newcommand{\gtsima}{$\; \buildrel > \over \sim \;$}
\newcommand{\gsim}{\lower.5ex\hbox{\gtsima}}
\newcommand{\bra}{\langle}
\newcommand{\ket}{\rangle}

\newcommand{\dd}{\mathrm{d}}

\newcommand{\likeli}{\mathcal{L}}
\newcommand{\ham}{\mathcal{H}}

\algdef{SE}[REPEAT]{Repeat}{EndRepeat}[1]{\textbf{repeat} #1 \textbf{times}}{}

\begin{document}

\title{Partition function approach to non-Gaussian likelihoods: \\ macrocanonical partitions and replicating Markov-chains \vspace{-30pt}}

\author{Maximilian Philipp Herzog\,\orcidlink{0009-0007-6635-496X}$^{2, \star, \diamond}$}
\author{Heinrich von Campe\,\orcidlink{0009-0006-4071-3576}$^{2, \star}$}
\author{Rebecca Maria Kuntz\,\orcidlink{0009-0006-0960-817X}$^{2}$}
\author{\\ Lennart R{\"o}ver\,\orcidlink{0000-0002-4248-8329}$^{1,2}$}
\author{Bj{\"o}rn Malte Sch{\"a}fer\,\orcidlink{0000-0002-9453-5772}$^{2, \sharp}$}
\thanks{$^\star$ Both authors contributed equally to this work.}
\thanks{$^\diamond$ \href{mailto:maximilian.herzog@stud.uni-heidelberg.de}{maximilian.herzog@stud.uni-heidelberg.de}}
\thanks{$^\sharp$ \href{mailto:bjoern.malte.schaefer@uni-heidelberg.de}{bjoern.malte.schaefer@uni-heidelberg.de}}

\affiliation{$^{1}$ Institut f{\"u}r Theoretische Physik, Universit{\"a}t Heidelberg, Philosophenweg 16, 69120 Heidelberg, Germany}
\affiliation{$^{2}$ Zentrum f{\"u}r Astronomie der Universit{\"a}t Heidelberg, Astronomisches Rechen-Institut, Philosophenweg 12, 69120 Heidelberg, Germany}

\begin{abstract}
Monte-Carlo techniques are standard numerical tools for exploring non-Gaussian and multivariate likelihoods. Many variants of the original Metropolis-Hastings algorithm have been proposed to increase the sampling efficiency. Motivated by Ensemble Monte Carlo we allow the number of Markov chains to vary by exchanging particles with a reservoir, controlled by a parameter analogous to a chemical potential $\mu$, which effectively establishes a random process that samples microstates from a macrocanonical instead of a canonical
ensemble. In this paper, we develop the theory of macrocanonical sampling for statistical inference on the basis of Bayesian macrocanonical partition functions, thereby bringing to light the relations between information-theoretical quantities and thermodynamic properties. Furthermore, we propose an algorithm for macrocanonical sampling, \texttt{Avalanche Sampling}, and apply it to various toy problems as well as the likelihood on the cosmological parameters $\Omega_m$ and $w$ on the basis of data from the supernova distance redshift relation.
\end{abstract}


\maketitle

\section{Introduction}
In problems of statistical inference, Bayes'\@ theorem combines the prior information $\pi(\theta)$ on the parameters~$\theta$ of a physical model with the likelihood $\likeli(y|\theta)$ as the distribution of the data points $y$ for a given parameter choice $\theta$ to the posterior distribution $p(\theta|y)$,
\begin{equation}
p(\theta|y) = \frac{\likeli(y|\theta)\pi(\theta)}{p(y)}
\end{equation}
with the Bayesian evidence
\begin{equation}
p(y) = \int\dd^n\theta\:\likeli(y|\theta)\pi(\theta)
\end{equation}
as the normalisation. The structure of Bayes'\@ theorem with an integral in the denominator and the integrand in the numerator suggests the definition of the canonical partition function
\begin{align} \label{eq:canZ}
&Z[\beta,J_\alpha] =
\int\dd^n\theta\:\left[\likeli(y|\theta)\pi(\beta)\:\exp(J_\alpha\theta^\alpha)\right]^\beta \\ &=
\int\dd^n\theta\:\exp\left(-\beta\left[\frac{1}{2}\chi^2(y|\theta)+\phi(\theta)\right]\right)\:\exp(\beta J_\alpha\theta^\alpha) \nonumber\\
&= \int\dd^n\theta\:\exp\left(-\beta\mathcal{H}(\theta)\right)\:\exp(\beta J_\alpha\theta^\alpha)
\end{align}
which falls back on the Bayesian evidence $p(y)$ for $\beta = 1$ and $J_\alpha = 0$. By differentiation of the partition function $\ln Z/\beta$ with respect to $J_\alpha$, cumulants of the posterior distribution $p(\theta|y)$ can be computed, making them easily accessible beyond second order.

As cosmological likelihoods generally have non-Gaus\-sian shapes and are joint distributions in a large number of parameters with possibly strong degeneracies between them, analytical methods are difficult to employ beyond the second order. As soon as a Gaussian approximation to the likelihood is permissible, the Fisher-formalism can play its strength \citep{tegmark_karhunen-loeve_1997}, and has found applications throughout cosmology \citep[for instance,][]{bassett_fisher4cast_2009, bassett_fisher_2011, coe_fisher_2009, elsner_fast_2012, refregier_icosmo:_2011, amara_figures_2011, raveri_cosmicfish_2016}. Extensions to the Fisher-matrix formalism including higher-order derivatives for dealing with non-Gaussian likelihoods have been constructed that are able to capture non-Gaussian shapes of distributions \citep{wolz_validity_2012, giesel_information_2021, schafer_describing_2016, sellentin_breaking_2014}.

Writing both likeli\-hood and prior in ex\-po\-nen\-tial form leads to a clear physical pic\-ture of how the Metro\-polis-Has\-tings algorithm samples tuples $\theta^\alpha$ as microstates from the posterior distribution $p(\theta|y)$: Effectively, \mbox{$\smash{\chi^2(y|\theta)/2+\phi(\theta)}$} is a potential, on which the Markov chain performs a thermal random walk at unit inverse temperature $\beta = 1$. Thus, the samples $\theta^\alpha$ appear with probabilities proportional to $\smash{\exp(-(\chi^2(y|\theta)/2+\phi(\theta)))}$, ensuring transitivity with a Boltzmann-type weighting -- samples with positive differences in
$\smash{\chi^2(y|\theta)/2+\phi(\theta)}$ are exponentially suppressed. Since its first introduction by \citet{lewis_cosmological_2002}, Markov chain Monte Carlo evaluation of posterior distributions has become a central tool in cosmology, and many variants of the original Metropolis-Hastings algorithm have been conceived to improve sampling efficiency.

Motivated by \citet{jaynes_information_1957}, who argues that thermodynamics is in its core a theory of information, we aim to use ideas from thermodynamics in statistical inference. Previously, \citet{10.1093/mnras/Rover} established analytical methods akin to a canonical Bayesian partition function in order to understand how the information contained in data reduces uncertainty on the parameter choice of a physical model.

Recently, inspired by the micro-canonical ensemble, other groups proposed Microcanonical Hamilton Monte Carlo \citep{micro1,micro2}, where the kinetic energy term is tuned such that the marginal over the uniform distribution over constant energy-shells is the target distribution. We believe that ensemble-inspired sampling is a fruitful approach.

To further explore these ideas we replace the canonical partition sum \eqref{eq:canZ} by a macrocanonical one, which should offer advantages in dealing with strongly degenerate and multimodal distributions.

Instead of controlling the number of independent samplers $N$ in a canonical ensemble, which would be described by the partition function $Z[\beta,J_\alpha,N] = Z[\beta,J_\alpha]^N$, one can carry out a
Laplace transform to replace the dependence on $N$ by a dependence on the chemical potential $\mu$, leading
to the macrocanonical partition function
\begin{align} \label{eq:xi}
\Xi[\beta,J_\alpha,\mu] &=
\sum_N\frac{1}{N!}Z[\beta,J_\alpha,N]\exp(\beta\mu N)\nonumber \\ &=
\sum_N\frac{1}{N!}\left(Z[\beta,J_\alpha]\exp(\beta\mu)\right)^N \nonumber \\ &=
\exp\left(zZ[\beta,J_\alpha]\right)
\end{align}
introducing the Gibbs-factor $1/N!$ and the fugacity \mbox{$z = \exp(\beta\mu)$}.

In statistical physics, the chemical potential signifies the energy required for adding a particle to the system. Controlling $\mu$ instead of $N$ suggests a number of advantages: This provides a natural way to control the sampling process in a manner reminiscent of simulated annealing, where the (inverse) temperature changes according to a schedule. In general, one would start with high temperatures such that the entirety of the parameter space is explored, before systematically reducing the temperature to $\beta = 1$. Additionally, when reaching thermal equilibrium, the number of samplers $N$ approaches a Poissonian distribution around $\langle N \rangle = p(y)e^{\mu}$ at unit temperature. But most importantly, one can argue that the sampling of multimodal distribution can be made more efficient: This is reached by the choice of a small value for $\mu$, such that the samplers can procreate close to a local minimum in $\mathcal{H}(\theta)=\chi^2(y|\theta)/2+\phi(\theta)$, leading to the enhanced generation of samples at favourable positions in parameter space.

Our paper is structured as follows: The theory of macrocanonical partition functions for statistical inference is summarised in Sect.~\ref{sect_partitions}. Our implementation of a macrocanonical sampling algorithm is presented in Sect.~\ref{sect_macrosampling} and applied to supernova data in Sect.~\ref{sec:supernovae}. We discuss our main results in Sect.~\ref{sect_summary}. Throughout the paper, we adopt the summation convention and denote parameter tuples $\theta^\alpha$ and data tuples $y^i$ as vectors with contravariant indices; Greek indices are reserved for quantities in parameter space and Latin indices for objects in data space. Concerning the cosmological example, we assume a flat, dark energy-dominated Friedmann-Lema{\^i}tre-Robertson-Walker spacetime with matter density $\Omega_m$ and a constant dark energy equation of state parameter $w$.

\section{Partition functions and Metropolis-Hastings sampling}\label{sect_partitions}

\subsection{Macrocanonical partition sums and derived quantities}
Previously, \cite{partitionfunction101} have shown that derivatives of the canonical partition sum \eqref{eq:canZ}, and the canonical potential
\begin{equation}
    F(\beta, J_\alpha) = -\frac{1}{\beta} \ln Z[\beta, J_\alpha]
\end{equation}
motivated by analogy to thermodynamics, allow us to find the moments
\begin{equation}
\langle \theta^{\mu_1} \ldots \theta^{\mu_m} \rangle_{\theta \sim p(\theta \mid y)}
= \frac{1}{Z} \partial^{\mu_1} \ldots \partial^{\mu_m} Z
\big|_{\substack{\beta=1,\\J=0}}
\end{equation}
as well as the cumulants
\begin{equation}
\langle \theta^{\mu_1} \ldots \theta^{\mu_m} \rangle^c_{\theta \sim p(\theta \mid y)}
= -\partial^{\mu_1} \ldots \partial^{\mu_m} F
\big|_{\substack{\beta=1,\\J=0}}
\end{equation}
of the posterior. Here, $\partial^\mu = \partial / \partial J_\mu$ denote derivatives w.\,r.\,t.\@ the sources. In analogy, we define the macrocanonical potential
\begin{equation}
    \Omega(\beta, J_\alpha, \mu) = -\frac{1}{\beta} \ln \Xi[\beta, J_\alpha, \mu] = -\frac{1}{\beta} z Z[\beta, J_\alpha],
\end{equation}
to calculate the expected number of Markov-chains in the macrocanonical ensemble
\[
\langle N \rangle
= - \frac{\partial \Omega}{\partial \mu}
\big|_{\substack{\beta=1\\J=0}}
= z Z
\big|_{\substack{\beta=1\\J=0}}
= e^\mu p(y).
\]
This result is quite interesting: The average number of chains is exactly the Bayesian evidence, up to a prefactor characterised by the chemical potential. Conversely, one could state that merely counting the chains is enough to recover the evidence. The distribution of the
number of chains will be discussed in detail in the next section.
Meanwhile, the expected
(total) energy of the macrocanonical ensemble is
\begin{align} \label{eq:Htot}
\langle \mathcal{H}_\text{tot} \rangle
&= \frac{1}{\Xi} \sum_N \frac{1}{N!} z^N \int \dd^{N\cdot n}{\theta} \sum_i \mathcal{H}(\theta_i) e^{-\beta \sum_i \mathcal{H}(\theta_i)} \nonumber\\
&= - \frac{1}{\Xi} \sum_N \frac{1}{N!} z^N \frac{\partial}{\partial \beta} Z[\beta]^N \nonumber\\
&= - \frac{1}{\Xi} \sum_N \frac{1}{N!} z^N N Z^N \frac{1}{Z} \frac{\partial Z}{\partial \beta} \nonumber\\
&= \langle N \rangle \langle \mathcal{H} \rangle_{N=1}.
\end{align}
This result is intuitive: The expected total energy of the system is the product of the expected number of chains and the expected energy for a single Markov-chain,
\mbox{$\langle \mathcal{H} \rangle_{N=1} = -Z^{-1} \partial_\beta Z$}.

Next, we would like to compute the moments and cumulants. Starting from the expression with the canonical partition sum, we find
\begin{align}\label{eq:cumulant}
\langle \theta^{\mu_1} \ldots \theta^{\mu_m} \rangle_{\theta \sim p(\theta \mid y)}
&= \frac{1}{Z} \partial^{\mu_1} \ldots \partial^{\mu_m} Z \big|_{\substack{\beta=1\\J=0}}\nonumber \\
&= \frac{1}{\langle N \rangle} \partial^{\mu_1} \ldots \partial^{\mu_m}
\langle N \rangle
\big|_{\substack{\beta=1\\J=0}},
\end{align}
using $z Z = \ln \Xi = \langle N \rangle$, and
\begin{align}
\langle \theta^{\mu_1} \ldots \theta^{\mu_m} \rangle^c_{\theta \sim p(\theta \mid y)}
&= \partial^{\mu_1} \ldots \partial^{\mu_m} \ln Z
\big|_{\substack{\beta=1\\J=0}} \nonumber \\
&= \partial^{\mu_1} \ldots \partial^{\mu_m} \ln \langle N \rangle
\big|_{\substack{\beta=1\\J=0}}.
\end{align}
These results are typical, especially the double appearance of the logarithm in the second equation is characteristic for calculations in the macrocanonical ensemble. Still, the omnipresence of the expected number of chains $\langle N \rangle$ replacing $Z$ as generating functional is remarkable.

Again considering \eqref{eq:cumulant}
\begin{align*}
\langle \theta^{\mu_1} \ldots \theta^{\mu_m} \rangle_{\theta \sim p(\theta \mid y)}
&= -\partial^{\mu_1} \ldots \partial^{\mu_m} \left(
\frac{1}{\langle N \rangle_{J=0}} \Omega \right)
_{\substack{\beta=1\\J=0}} \\
&= \frac{1}{Z} \partial^{\mu_1} \ldots \partial^{\mu_m} Z
\big|_{\substack{\beta=1\\J=0}},
\end{align*}
we recover the same result when dividing out the expected number of particles $\langle N \rangle_{J=0}$ before the derivatives.

The Cram{\'e}r-Rao inequality states that the (average) parameter covariance given by a posterior is bounded from below by the (average) inverse Fisher information, $\langle \theta^\mu \theta^\nu \rangle^c \geq F^{\mu \nu}$. \citet{partitionfunction101} have shown that, starting from the canonical partition function for a Gaussian distribution,
\begin{equation} \label{eq:Zgauss}
Z [\beta, J_\alpha]
= \sqrt{\frac{(2 \pi)^n}{\beta^n \det F}} \exp\left(\frac{\beta}{2} F^{\mu \nu} J_\mu J_\nu\right),
\end{equation}
one indeed recovers this lower bound. Since we have found a way to compute the same second cumulant within the framework of the macrocanonical ensemble, the result holds completely analogously here,
\[
\langle \theta^\mu \theta^\nu \rangle^c_{\theta \sim p(\theta | y)}
= \partial^\mu \partial^\nu \ln \ln \Xi
\big|_{\substack{\beta=1\\J=0}}
= F^{\mu \nu}.
\]

It might be worthwhile to compare macrocanonical sampling to affine-invariant sampling methods such as \texttt{emcee}: There, on achieves in cases of strong statistical degeneracies an efficiency boost by affine scaling of the parameter space such that the proposal distribution can be chosen to be symmetric. In the language of the canonical partition sum $Z[\beta,J_\alpha]$ for e.g. a quadratic $\chi^2(y|\theta)$-functional and disregarding a prior $\pi(\theta)$,
\begin{equation}
Z[\beta,J_\alpha] =
\int\dd^n\theta\:\exp\left(-\beta\left[F_{\alpha\beta}\theta^\alpha\theta^\beta + J_\alpha\theta^\alpha\right]\right),
\end{equation}
one immediately notices affine invariance of the two terms $F_{\alpha\beta}\theta^\alpha\theta^\beta$ and $J_\alpha\theta^\alpha$ as proper scalar quantities: The linear forms $F_{\alpha\beta}$ and $J_\alpha$ transform inversely to $\theta^\alpha$. The volume element, $\dd^n\theta$, however, would only be invariant if the covolume is included, $\dd^n\theta\sqrt{\det F}$, but as the normalisation of partition functions are not fixed but rather generated in the logarithmic derivatives, this particular factor does not play a role.

Transitioning from canonical to macrocanonical ensembles implies the replacement of particle number $N$ by the chemical potential $\mu$ as a state variable. Along this process, the question arises whether this eventually leads to a conflict with the Gibbs-Duhem relation, which states that at least one state variable needs to be extensive such that information about the physical size of the system is retained. We would argue that the ensemble of samplers considered here do not correspond to a physical system, and that extensivity of quantities like e.g. the equivalent total energy (see eqn.~\ref{eq:Htot}) is not a necessity.

Combining eqn.~\ref{eq:xi} for the macrocanonical partition with the analytical expression eqn.~\ref{eq:Zgauss} for the Gaussian canonical parition \citep{partitionfunction101, kuntz2023partition} leads to the analytical expression
\begin{equation}
\Xi[\beta,J_\alpha,\mu] = 
\exp\left(z\sqrt{\frac{(2\pi)^n}{\beta^n\det F}}\exp\left(\frac{\beta}{2}F^{\mu\nu}J_\mu J_\nu\right)\right)
\end{equation}
with the typical nested exponential.

\subsection{Distribution of the number of chains} \label{sec:distOfN}
The expression of the macrocanonical partition sum~(\ref{eq:xi}) gives insight into the expected number of chains in a macrocanonical MCMC algorithm as well as the distribution of the number of chains in thermodynamic
equilibrium. In fact, the distribution of the number $N$ of particles associated to $\Xi[\beta,J_\alpha,\mu]$ can be expressed as
\begin{align}
    \label{eq:poissonian}
	p(N|\beta, \mu) &=
	\frac{1}{\Xi[\beta, \mu]}\frac{1}{N!} \left(\exp(\beta \mu) Z[\beta]\right)^N \nonumber \\ &=
	\frac{1}{N!} \left(\exp(\beta \mu) Z[\beta]\right)^N \exp\left(-\exp(\beta \mu) Z[\beta]\right),
\end{align}
which is effectively a Poisson distribution with expectation value and all higher cumulants equal to \mbox{$\langle N
\rangle = \exp(\beta \mu) Z[\beta]$}. In a
macrocanonical sampler $\mu$ determines and controls the number of
Markov-chains $N$ in thermodynamic equilibrium. Therefore, being able to compute $N$ lets us diagnose the sampling process and ascertain if it has reached thermodynamic equilibrium.
\footnote{Naturally, if we know $Z[\beta]$ we may control the distribution of $N$ directly by choosing $\mu$ accordingly. Note that the value obtained this way can easily be negative, if $Z[\beta]$ is smaller than the desired average amount of particles.}

One noteworthy remark is that the evidence, which arises as $Z[\beta=1] = p(y)$, can be recovered from the sampler via $\langle N \rangle_{\beta=1} = e^\mu p(y)$ by counting the number of chains. There is a great variety of algorithms designed to recover the evidence, since the log model evidence $\ln p(y|M)$ is a crucial tool for model comparison. One such algorithm is Nested Sampling \citep{Skilling_2006, Ashton_2022, Buchner_2023}, which iteratively partitions the parameter space into nested subsets with progressively higher likelihoods before computing $p(y)$ as a likelihood-weighted sum of the covered prior volumes. Note that Nested Sampling also provides an estimate of the posterior distribution as a by-product.

Despite its great popularity, Nested Sampling does suffer from several shortcomings, such as the introduction of additional noise when probing the prior by scattering initial particles \citep[see discussion by][]{dynesty_2020}. Additionally, the prior has to be normalisable. Conversely, in merging the notion of the macrocanonical ensemble with fact that $\ln p(y) = \ln \langle N \rangle_{\beta=1} - \mu$, we successfully constructed a physically motivated procedure for estimating the evidence. Crucially, this method is free of said noise. Instead, as the statistics of $\ln p(y)$ follows the Poissonian statistics of $N$, the variance is just given by
\begin{equation*}
    \Delta \ln p(y) = \frac{\Delta \langle N \rangle}{\langle N \rangle} = \frac{1}{\langle N \rangle} \frac{\Delta N}{k} = \frac{1}{k},
\end{equation*}
where $k$ is the number of samples used to estimate $\langle N \rangle$. Also, it does not require a normalisable prior. Simultaneously, it also provides a good estimate of the posterior.
\footnote{Indeed, the fact that we do not require the prior to be normalisable makes this approach able to universally estimate integrals over any non-compact subspace $U\subseteq \mathbb{R}^n$ if the integrand is $L^1(U)$.}

Finally, changes in $\mu$ are a way of interacting with the Markov chains, in the same way as introducing slow changes in $\beta$ during the sampling process. Such manipulations allow moving from a highly-populated exploratory mode to a mode where a few chains focus on deep minima, making the sampling more efficient. As a pendant to simulated annealing, we jokingly refer to this as simulated annihilation.

\subsection{Shannon's entropy and heat capacity}
In the context of the Bayesian canonical partition sum~\eqref{eq:canZ}, \citet{partitionfunction101} have demonstrated that with
\begin{align}
\frac{\partial Z}{\partial \beta}
&= \frac{\partial}{\partial \beta} \int \dd^n{\theta}\: [\mathcal{L} \pi]^\beta = \int \dd^n{\theta}\: [\mathcal{L} \pi]^\beta \ln (\mathcal{L} \pi)\nonumber \\
&= \int \dd^n{\theta}\: [p(\theta \mid y) p(y)]^\beta
\ln (p(\theta \mid y) p(y))
\end{align}
one may derive the differential Shannon entropy of the posterior as a derivative of the free energy
$F(\beta,J_\alpha) = -\ln Z[\beta,J_\alpha]/\beta$ at unit temperature,
\begin{align*}
S &= \left. \beta^2 \frac{\partial F}{\partial \beta}
\right|_{\beta=1}
= \beta^2 \frac{\partial}{\partial \beta} \left(-\frac{1}{\beta} \ln Z\right)_{\beta=1} \nonumber \\
&= -\int \dd^n{\theta}\: p(\theta \mid y) \ln p(\theta \mid y)
= - \left\langle \ln p(\theta \mid y) \right\rangle_{\theta \sim p(\theta | y)}.
\end{align*}
Going one step further, they also computed the heat capacity as the second derivative of the free energy,
\begin{align}
C
&= \beta \frac{\partial}{\partial \beta}\left[-\beta^2 \frac{\partial F}{\partial \beta}\right]_{\beta = 1}
= \left\langle \mathcal{H}^2 \right\rangle^c_{\beta=1,\: \theta \sim p(\theta | y)} \nonumber\\
&= \left\langle \left(-\chi^2(y \mid \theta) / 2
- \phi(\theta) \right)^2 \right\rangle^c_{\theta \sim p(\theta | y)}.
\end{align}

While it is a well known result that the heat capacity of the canonical ensemble equals the variance of the energy, the last expression is noteworthy: this term does not have a clear interpretation in the context of information theory. It appears to be an emphasised measure of the dispersion in $\likeli \pi$.
In the following, we aim to perform analogous calculations for the macrocanonical partition sum. Omitting the sources $J_\alpha$ and recalling the grand canonical potential $\Omega = -z Z/\beta$, we begin with the entropy,
\begin{align*}
&S_\text{macro} = \left. \beta^2 \frac{\partial \Omega}{\partial \beta} \right|_{T=1} = \beta^2 \frac{\partial}{\partial \beta} \left(-\frac{1}{\beta} z Z\right)_{\beta=1}\nonumber \\
&= e^\mu p(y) \left[ 1 - \mu - \ln p(y) - \int \dd^n{\theta} p(\theta \mid y) \ln p(\theta \mid y)\right].
\end{align*}
Keeping in mind that
$\langle N \rangle_{\beta=1} = e^\mu p(y)$, we get
\begin{equation}\label{eq:S of xi}
S_\text{macro} = \langle N \rangle - \langle N \rangle \ln \langle N \rangle + \langle N \rangle\; S
\end{equation}
The last term is in accordance with our expectations, as it is constituted by the product of Shannon's entropy of the posterior and the expected number of particles. The fact that the entropy depends solely on $\langle N \rangle$ is anticipated, as the number of chains should be sufficient to quantify the amount of information
stored in the sampler positions.

Indeed, one may derive the same result entirely on the thermodynamic consequence of indistinguishability: For
the \emph{microcanonical} partition sum $W[E,N]$, the entropy is defined as
$S = \ln W$(setting Boltzmann's constant $k_\text{B}=1$). Now, if we consider
the case of $N$ particles ($N$ fixed) instead and replace $W$ by $W^N/N!$, we find
\begin{equation*}
\ln \frac{1}{N!} W^N = -\ln N! + N \ln W
\approx N  -N \ln N + N \cdot S,
\end{equation*}
which is the same result as derived in \eqref{eq:S of xi} with $\langle N \rangle$.

It is interesting to note that the derived entropy is non-extensive, and it may take negative values. On the one hand, this is due to Shannon's entropy only being a strictly positive functional on a discrete support, so $S$ by itself can be negative if we consider a continuous parameter space. On the other hand, the indistinguishability of the particles reduces the entropy by introducing $\ln N!$ nats of information.
For $N \gg 1$ the particles begin to look alike, such that the factor $N!$ is over counting the number of distinguishable states and the system becomes over determined.

For the heat capacity we find
\begin{align}
C_\text{macro} &= \beta \frac{\partial}{\partial \beta}\left[-\beta^2 \frac{\partial \Omega}{\partial \beta}\right]_{\beta = 1} \nonumber\\
&= \langle N \rangle\; C + \langle N \rangle \left(\ln \langle N \rangle - S \right)^2.
\end{align}
It is worth noting that the heat capacity of the canonical ensemble $C$ is recovered in this expression alongside a linear factor of $\langle N\rangle$. In accordance with our expectations, the heat capacity does not depend on $\mu$.
We attribute the positive adjustment term in $C_\text{macro}$, to the ability of the system to gain more heat by exchaning particles with the particle reservoir.

\subsection{Equipartition} \label{subsec:equipartition}
A properly equilibrated Markov chain fulfils equipartition conditions \citep{10.1093/mnras/Rover}. This stipulates that the degrees of freedom of the system separate in the sense that
\begin{equation}
\left\langle \theta^\mu\frac{\partial\mathcal{H}}{\partial \theta^\nu}\right\rangle = \frac{1}{\beta} \delta^\mu_\nu.
\end{equation}
These relations follow naturally from the macrocanonical partition function $\Xi[\beta,J_\alpha,\mu]$. Under the assumption of a constrained potential we find
\begin{align} \label{eq:equipartition}
\left\langle \theta^\mu\frac{\partial\mathcal{H}}{\partial\theta^\nu}\right\rangle &=
\frac{1}{\beta\Xi}\sum_N \frac{1}{N!}\left(z \int \dd^{n}\theta\:\frac{\partial\theta^\mu}{\partial\theta^\nu}e^{-\beta \mathcal{H}(\theta)}\right)^N \nonumber \\&= \frac{1}{\beta}\delta^\mu_\nu,
\end{align}
which again yields diagnostic of chain convergence.

\subsection{Markov-processes and detailed balance} \label{sec:detailedBalance}
From an abstract, probabilistic perspective, the movement of a particle in a Markov process \citep{metropolis_equation_1953, Hastings_1970, metropolis_monte-carlo:_1985} corresponds to a series of states
of some system where the current state may depend on nothing but the last state.
Apart from that, we want the transition graph between states to fulfill \emph{ergodicity} and \emph{irreducibility}, i.\,e.\@ that any point in configuration space will be reached with certainty in a finite number of steps.
Further, it should adhere to the condition of \emph{detailed balance}. This can be viewed as the macroscopic articulation of the equilibrium principle of microscopic time reversibility. Concretely, the principle of detailed balance is manifested in the choice for the probability $\alpha$ of \emph{accepting} a transition from $\theta \to \theta'$. Given a \emph{proposal} distribution $\theta' \sim T(\theta' | \theta)$ and a stationary target distribution $p^*(\theta\;)$, a suitable acceptance probability $\alpha$ is given by 
\begin{equation} \label{eq:alpha}
    \alpha(\theta', \theta) = \min\left\{1, \frac{T(\theta\; \mid \theta')p^*(\theta')}{T(\theta' \mid \theta\;)p^*(\theta\;)} \right\}. 
\end{equation}
The reasoning behind this well-known result can be found in the Appendix. 

\section{A Macrocanonical sampling algorithm:\\\texttt{Avalanche Sampling}}
\label{sect_macrosampling}
We believe that the use of a macrocanonical sampling procedure is a promising approach to multimodal cosmological inference problems. Such an algorithm is able to continuously create and kill samplers to the benefit of the overall sampling efficiency and accuracy in these situations. As opposed to other algorithms
such as \textit{Ensemble Monte Carlo} \citep[for instance, ][]{emcee}, macrocanonical samplers are able to dynamically adapt to the requirements of the underlying potential, making them especially well-suited for effective sampling of complex structures.

\begin{figure}
    \centering
    \includegraphics*[width=0.47\textwidth]{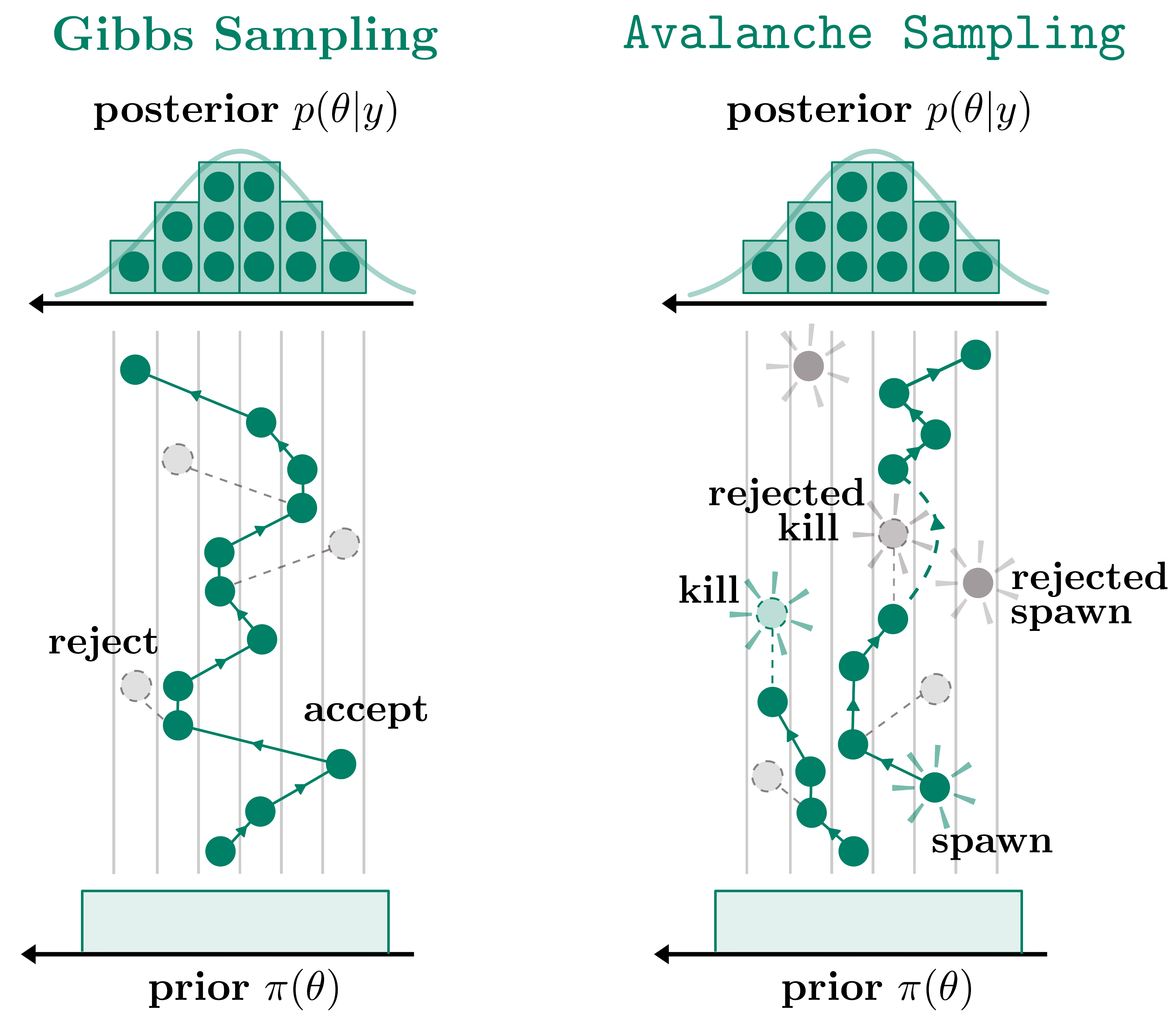}
    \label{fig:illustration_gibbs_vs_avalanche}
    \caption{Schematic comparison of Gibbs sampling and our proposed \texttt{Avalanche Sampling}. In both schemes, accept and reject correspond to green and grey fillings respectively. Different kinds of update steps include \enquote{kill/spawn} (as indicated by grey circles and radial lines respectively) and \enquote{move} (as indicated by arrows) \cite[Illustration inspired by][Figure 2]{MCMC_plot}}
\end{figure}

\subsection{The algorithm}
We propose the generation of Metropolis-Hastings samples in two steps where we differentiate between \enquote{moving} and \enquote{kill/spawn} updates. This separation is possible as the particle position
statistics are independent of the statistics of the particle quantity and vice versa.

From the grand canonical partition function~\eqref{eq:xi}, we may identify the target distribution
\begin{align} \label{eq:pStar}
p^*(\{\theta_1\dots\theta_N\})
&= \frac{1}{\Xi}\: e^{\beta \mu N - \beta\sum_{i=1}^N \ham(\theta_i)}\nonumber \\
&= \frac{1}{\Xi} e^{\beta \mu N} \prod_{i=1}^N e^{-\ham(\theta_i)}.
\end{align}
This factorisation demonstrates how particles may move individually without
affecting each other. For each move, we choose a symmetric proposal distribution $T(\theta_i' \mid \theta_i)$ (e.\,g.\@ a Gaussian) only depending on the position of the particle in question.

Since it is symmetric, it will drop out in \eqref{eq:alpha}. And because of the factorisation, also all terms concerning the other particles will cancel, yielding
\[
\alpha(\theta_i', \theta_i) = \min \left\{ 1, e^{ \
- \ham(\theta_i') + \ham(\theta_i) \\
}\right\}.
\]

\noindent
Of course, this first \enquote{moving} step allows for a lot of variation, using modifications that have been proposed in order to improve the performance standard Metropolis Hastings sampling. An especially
interesting example is the case of Hamilton Monte Carlo \citep{HMC_2011}. While such modifications are in principle applicable to \texttt{Avalanche Sampling}, we limit ourselves to the vanilla case for simplicity.
Therefore, our algorithm is intended as an \enquote{add-on} to wrap around a more frequently used algorithm rather than a standalone procedure. Naming the method \texttt{Avalanche Sampling} was chosen because it reflects the behaviour when descending into a minimum of $\smash{\chi^2(y|\theta)/2+\phi(\theta)}$, when potential energy is traded for adding additional samplers to the system, which can recreate this process and ultimately form an avalanche rolling toward the minimum.

For the kill/spawn steps, this decoupling is also beneficial since it only requires the position of the particle in question as well as the number of particles The basic idea of the algorithm will thus be the
following: Mostly, particles move around as  described previously but every few move steps, we perform a kill/spawn step where we toss a coin, deciding whether to attempt the creation or annihilation of a particle. But how does one choose the proposal distributions for spawning and killing particles? We have devised two \emph{modes} for this:

\vspace{5pt}\noindent
\texttt{static spawn} ---
As an initial idea, the particles spawn according to a proposal distribution
that does not depend on the current state,
\begin{equation} \label{eq:proposeSpawnStatic}
T(^*\theta_{N + 1}) = \frac{1}{2} \:p(\theta_{N + 1}).
\end{equation}
Here $\smash{T(^*\theta_{N+1}) := T(\{\theta_1,\dots , \theta_N, \theta_{N + 1}\}| \{\theta_1 ,\dots, \theta_N \})}$ is the proposal distribution for spawning a particle at $\theta_{N+1}$, given the positions of all other particles. The factor $\nicefrac{1}{2}$ stems from the \enquote{coin toss} deciding about the creation/annihilation of a particle.
Meanwhile, the simplest way to choose a particle to kill is just at random, so the distribution for proposing killing the particle at index k, $T(
^\dagger \theta_k) := T(\{\theta_1,  \dots,\theta_{k-1} ,\theta_{k+1} ,\dots , \theta_N\}| \{\theta_1 ,\dots ,\theta_k, \dots, \theta_N \})$, is
\begin{equation} \label{eq:propseKillStatic}
T(^\dagger \theta_k) = \frac{1}{2N}.
\end{equation}

\begin{figure*}[t]
    \begin{minipage}{\linewidth}
    \begin{algorithm}[H]
    \caption{\texttt{Avalanche Sampling}}
    \label{algo:macromcmc}
    \begin{algorithmic}[0]
    \State \textbf{input:} Likelihood $\mathcal{L}(y | \theta) = \exp[- \frac{1}{2} \chi^2(y | \theta)]$, prior $\pi (\theta) = \exp[-\phi(\theta)]$, $\ham=\chi^2/2 + \phi$
    \State \hspace{22pt} Chemical potential $\mu$,
    static spawn proposal distribution $p(\theta)$
    \textbf{or} \enquote{closeness} kernel $p(\theta_i \mid \theta_j)$,
    $N_{\text{steps}}$ and initial number of chains
    $N_\text{init}$
    \State \hspace{22pt} \textcolor{darkgray}{MCMC-parameters: symmetric \enquote{move} proposal distribution $T(\theta | \theta')$,
    $N_{\text{generations}}$ (as well as $N_{\text{burn-in}}$ and $N_{\text{thinning}}$  for data de-correlation)}
    \State \textbf{output:} Particle position samples $\Theta$
    \State
    \State \textit{Initialise particle positions $\{\theta^{j}_0\}_{j=1,\dots, N_{\text{init}}} \sim \pi(\theta)$}
    \Repeat{$N_{\text{generations}}$}
        \Repeat{$N_{\text{steps}}$} \Comment{Move $N_\text{step}$ particles \textcolor{darkgray}{(classical MCMC)}}
            \State \textcolor{darkgray}{\textit{Choose particle} $k \sim \mathcal{U}\{1, N\}$} \vspace{3pt}
            \State \textcolor{darkgray}{\textit{Propose new position} $\theta_k' \sim T(\cdot \mid \theta_k)$}
            \State \textcolor{darkgray}{\textit{Accept} $\theta_k \gets \theta_k'$
            \textit{with probability}
            $\alpha(\theta_k', \theta_k) = \min\left(1, \exp[-\ham(\theta'_k)+\ham(\theta_k)]\right)$}
        \EndRepeat
        \State \textit{Flip a coin} $r \sim \mathcal{U}\{0,1\}$ \Comment{Macrocanonical step}
        \If{\textit{Heads} $r = 0$}
            \State \textit{Propose a new particle}
            $\theta_{N+1} \sim T(^* \theta_{N+1})$
            \State \textit{Spawn particle} $\theta_{N+1}\;(N \to N+1)$
            \textit{with probability}
            $\alpha(^* \theta_{N+1})$
        \EndIf
        \If{\textit{Tails} $r = 1$}
            \State \textit{Choose particle}
            $k \sim T(^\dagger \theta_k)$
            \State \textit{Kill particle}
            $\theta_{k}\;(N \to N-1)$ \textit{with probability}
            $\alpha(^\dagger \theta_k)$
        \EndIf
    \EndRepeat
    \vspace{-5pt}
    \end{algorithmic}
    \end{algorithm}
    \end{minipage}
\end{figure*}

\vspace{5pt}\noindent
\texttt{proximity spawn} ---
In a more elaborate scheme, we want new particles to spawn near existing particles in order to explore a given minimum of the potential as well as possible. This is achieved in the Metropolis-Hastings algorithm,
as the particle will mostly remain close to a minimum. If particles \enquote{preferred} to spawn near other particles, we would find a higher number in the vicinity of minima. As opposed to the classical Metropolis-Hastings algorithm, our method is able to work with multiple minima, i.\,e.\@ multimodal distributions.

To construct the proposal probabilities, we require a probability density function $p(\theta_i \mid \theta_j)$ that expresses how \enquote{close} any pair of particles at positions $\theta_i, \, \theta_j$ is. It is computationally wise to make this symmetric in its two arguments, we chose a Gaussian for simplicity.
Thereupon, we may formulate the spawn proposal probability,
\begin{equation} \label{eq:proposeSpawnProximity}
T(^* \theta_{N +1})
= \frac{1}{2N} \sum_{i=1}^N p(\theta_{N+1} \mid \theta_i).
\end{equation}
Regarding the notion of detailed balance, it is intuitive that a particle is also more likely to be annihilated if it is close to other particles,
\begin{align} \label{eq:proposeKillProximity}
T(^\dagger \theta_k) = \frac{1}{2} \Bigg[ \sum_{i=1}^N \sum_{\substack{j=1\\j\neq i}}^N p(\theta_i \mid \theta_j) \Bigg]^{-1} \cdot
\sum_{\substack{l=1\\l \neq k}}^N p(\theta_l \mid \theta_k).
\end{align}
Note, that for $p(\theta_i \mid \theta_j) = p(\theta_i)$ we recover the \texttt{static spawn} proposal distributions \eqref{eq:proposeSpawnStatic} and \eqref{eq:propseKillStatic}, as this is a generalisation of the previous mode.

\vspace*{10pt}
For both modes, the acceptance probabilities follow from \eqref{eq:alpha}, the detailed expressions may be found in the appendix. With these in mind, the implementation of the macrocanonical extension to the
Metropolis-Hastings algorithm is outlined in pseudocode in \autoref{algo:macromcmc}.

Keep in mind that, since both modes fulfill detailed balance, which mode to chose is only a matter of convergence efficiency. The target distribution is the same and one can recover the evidence from the number of chains in both cases.
Still, the most advantageous choice of mode depends on the problem at hand: \texttt{static spawn} is very stable, 
no matter the number of dimensions $n$ and requires fewer computations. It has one disadvantage: Whenever a new chain spawns, it first needs to reach the closest minimum before its samples become usable. When the chains have only a short lifespan, this is not possible. Here, \texttt{proximity spawn} is clearly better since the new chains spawn already in favourable positions. A further discussion can be found in Sect.~\ref{numerical experiments}.
Also note that \texttt{proximity spawn} does not lead to oversampling of minima: 
One could intuitively think that since the algorithm only spawns chains around existing chains, the chains as a whole while never explore tails of the distribution. But, since the kill and spawn acceptance probabilities have to fulfill detailed balance, the chance of killing a particle is higher if it is surrounded by other particles. As this is naturally the case for minima, the chains will not oversample and will indeed explore the tails of the distribution.
For both algorithms, the choice of spawn kill proposal distributions has to be tuned. If it is too narrow, the chains will get stuck around local minima, never able to explore tails in the potential. If it is too wide, the acceptance rate will drop, making the algorithm inefficient. 
Difficulties with choosing the width of the kill and spawn proposal distributions in high-dimensional parameter spaces will also be discussed in Sect.~\ref{numerical experiments}.

In order to adhere to the Markov-property, we need to worry about whether subsequent samples are correlated and if the samples truly represent the target distribution.
A common approach is to exclude the commencing burn-in period by discarding the first samples such that only samples in an equilibrated state are regarded. We have various options to choose from, when it comes to diagnosing the convergence of our algorithm, see Sect.~\ref{subsec:equipartition} and \cite{10.1093/mnras/Rover}.
To avoid correlation between samples, we can throw away every $n$-th sample in a process called \emph{thinning}.
Still, for our algorithm, we need to take extra care: If we were to save all chains at every step, even those we did not touch, a sample at $i$ would not at all be correlated to a sample at $i+1$ but instead heavily correlated to a sample at $i+\langle N \rangle$, $i+2\langle N \rangle$, and so forth.
Instead, we only save the chains we proposed to move, spawn or kill, and if there is an accepted kill, or a rejected spawn, we choose a random chain to save its position. Interestingly, again, the correlation grows as the cummulative probability for choosing the same chain again increases.
After reaching a maximum, it drops finally, as information about its previous position is lost in random movement.

\subsection{Parallelisability}
A major advantage of Ensemble Monte Carlo is its parallelisability. It can be easily implemented on a GPU, where each thread holds a chain, cutting down on computation time. It has become necessary for any aspiring sampling algorithm to be parallelisable to some degree. For Nested Sampling there are also multiple ideas to parallelise the algorithm \citep[c.f. discussion by][]{Buchner_2023}.

For our proposed algorithm, this is not as straightforward, since we cannot predict the required number of threads. Nevertheless, we can think of some ways to parallelise the algorithm. For \texttt{static spawn}, chains can run simultaneously. After a specified time, we compute the likelihood of each particle, in order to perform a kill/spawn step where we either initialise a new thread or terminate one with respective acceptance probabilities.
The amount of chains $N$ can be shared as a global variable. For \texttt{proximity spawn}, we could look up to the methods applied in fluid simulations. Particles far away do not contribute significantly to the sums appearing in \eqref{eq:proposeSpawnProximity}. We can partition them into a grid and only compute the sums for particles in the same box and neighbouring boxes in a parallel fashion \citep{Green_2012}. This would again allow for a parallelisation of the kill/spawn steps. Clearly, further work on this is needed.

\section{Numerical Experiments}\label{numerical experiments}

\subsection{GAUSSIAN}
\paragraph*{Setup.} A Gaussian likelihood $\mathcal{N}(0,1)$ is employed as a toy example to test the algorithm, with
$\chi^2(y \mid \theta) = \theta_\alpha \theta^\alpha$
and an uninformative, uniform prior, i.\,e.\@ $\phi(\theta) \equiv 0$.

\begin{figure*}
    \centering
    \includegraphics[scale=0.97]{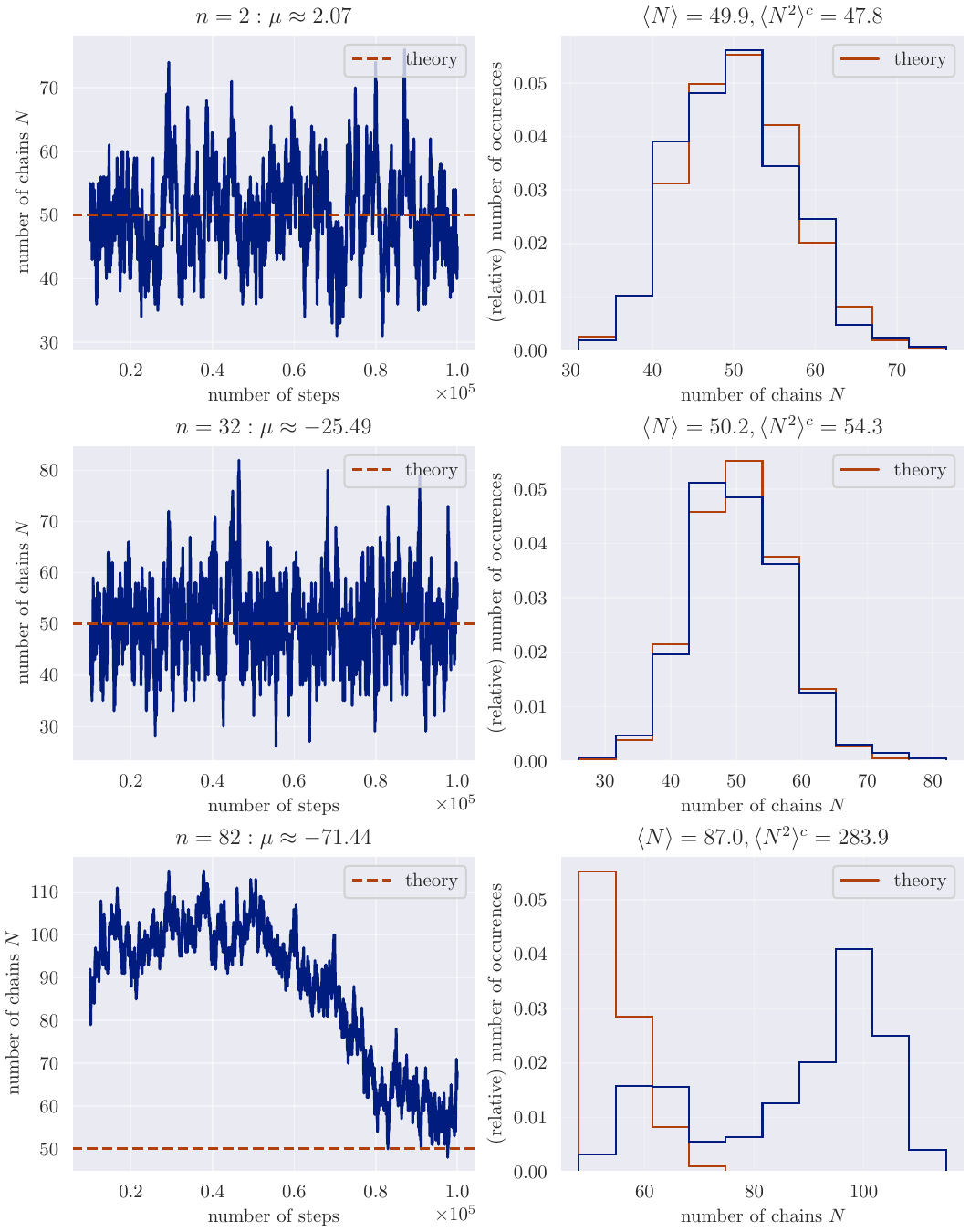}

    \caption{\textbf{$\mathbf{n}$-dimensional Gaussian} --- Left: The number of chains $N$
    as a function of Monte Carlo time;
    Right: A histogram of the number of chains. The distribution is Poissonian
    throughout the dimensions up to about $n\sim50$ and fails to be in higher dimensions:
    While the chemical potential $\mu$ was tuned using the analytical evidence given in \eqref{eq:Zgauss} such that
    $\langle N \rangle = 50$, this cannot be reproduced, as the acceptance rate is too low.
    This is due to the fact that for our algorithm it is not possible scale the spawn proposal distribution with the dimensionality accordingly.
    }
    \label{fig:n_chains}
\end{figure*}

\paragraph*{Results.}
First and foremost, our algorithm reproduces the expected Gaussian posterior distribution well, even in higher dimensions (c.f.~Fig.~\ref{fig:n_chains}).

Apart from that, this straightforward example serves as an effective testbed to verify whether we can accurately reproduce the Poissonian distribution for the number of chains as described in \eqref{eq:poissonian}.
For the Gaussian, the evidence may be calculated analytically according to \eqref{eq:Zgauss},
\begin{equation*}
    Z[\beta=1, J_\alpha=0] = p(y) = (2 \pi)^{n/2} \rightarrow \langle N \rangle = (2 \pi)^{n/2} e^\mu.
\end{equation*}

\begin{figure}[htp!]
    \centering
    \includegraphics{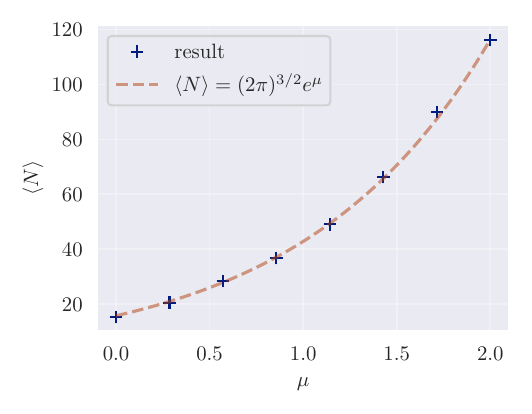}
    \caption{\textbf{$\mathbf{3}$-dimensional Gaussian} --- Average amount of chains $\langle N \rangle$ vs. chemical potential $\mu$. We can extract the evidence \mbox{$p(y) = Z[\beta=1]$} by counting the number of chains and finding the constant factor \mbox{$\smash{p(y) = \langle N \rangle e^{-\mu}}$}. For the 3-dimensional Gaussian, the analytical evidence is given by \mbox{$\smash{p(y) = (2 \pi)^{3/2}}$}.}
    \label{fig:extractZ}
\end{figure}

\begin{figure}
    \centering
    \includegraphics[scale=1]{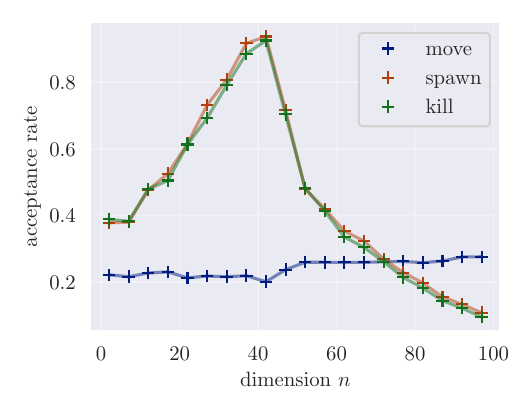}
    \caption{\textbf{$\mathbf{n}$-dimensional Gaussian} --- Acceptance rates for the \textit{static spawn} mode as a function of the number of dimensions $n$.
    While the acceptance rate for the move step stays mostly constant throughout increasing dimensions, the acceptance rate for spawn and kill first increase and then drop to $0$. Reasons for why we did not yet find a way of approaching such optimal scaling are given in the Sect.~\ref{numerical experiments}.
    }
    \label{fig:n_vsacceptanceProbabilities}
\end{figure}

Fig.~\ref{fig:extractZ} shows the exponential growth of $\langle N\rangle$ with increasing $\mu$. We succeed in reproducing the evidence. In Fig.~\ref{fig:n_chains}, it is evident that $N$ follows the corresponding Poissonian. Again, the algorithm performs quite well, however we observe a higher degree fluctuation in higher dimension, as given in the left part of the figure. This is due to the fact that the acceptance rate for spawn and kill steps decreases with increasing dimensionality, as illustrated in  Fig.~\ref{fig:n_vsacceptanceProbabilities}.
We will clarify this effect in the next few paragraphs.
Allowing more sampling steps would lead to equilibration of the algorithm at $N\simeq 50$ and a better approximation of Poisson-distribution after a longer time, as the trend in Fig.~\ref{fig:extractZ} suggests.

In fact this raises the concern of optimal scaling -- how to scale the covariance matrices with respect to the dimension, such that the acceptance rate stays optimal. \citet{Gelman_1992_optimal_scaling} have shown, that for a Metropolis-Hastings algorithm with a Gaussian random walk proposal, the variance of proposal distribution should scale with $\smash{n^{-1/2}}$, with a constant factor in front to achieve a constant optimal acceptance rate of about $23.4\%$. For the \textit{Metropolis-adjusted Langevin algorithm} (MALA), it is  $\smash{n^{-1/6}}$ \citep{Roberts_1998_optimal_scaling}, and for HMC the step-size of the leapfrog integrator should scale with $\smash{n^{-1/4}}$ \citep{HMC_optimal_scaling}. For our proposed algorithm, we have not yet found an analogous result. Although we found the most well-behaved acceptance rates for a scaling of $\smash{n^{-1/4}}$,
adjusting the covariance matrices cannot account for the offset induced by the changing chemical potential, so the acceptance rate necessarily drops to zero for large~$n$. In Fig.~\ref{fig:n_vsacceptanceProbabilities} we show acceptance rate for all three types of proposals in an $n$-dimensional Gaussian setting. While scaling the spawn covariance matrix with $\smash{n^{-1/4}}$, the chemical potential decreases linearly $\mu \propto -n$ to keep the average amount of chains constant.
If the algorithm converged, the acceptance rates for spawn and kill should be equal.
Due to the afformentioned reasoning, we see a tipping point for $n\sim40$, where the chemical potential term in the acceptance probability begins to dominate.
This is a major drawback of the algorithm, which we hope to address in future work.

It is also interesting to look at the average energy of the ensemble, at it incorporates both the number, and the movement of the chains individually. Using \eqref{eq:Htot} and \eqref{eq:Zgauss}, we get
\begin{align}
\langle \mathcal{H}_\text{tot} \rangle
&= -\langle N \rangle \frac{\partial}{\partial \beta} \ln Z[\beta] \Big|_{\beta=1} = \langle N \rangle \frac{n}{2}.
\end{align}

\begin{figure}
    
    \centering
    \includegraphics{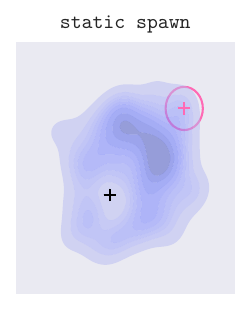}
    \includegraphics{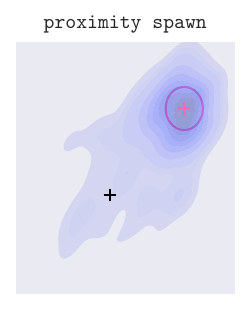}
    \caption{\textbf{$\mathbf{2}$-dimensional Gaussian} --- Result of a short run of \texttt{Avalanche Sampling} for a shifted 2-dimensional Gaussian potential $\mathcal{N}((4,4), I_2)$ \textcolor{hotpink}{\textbf{(pink)}}. \texttt{Static spawn} is not able to reach the minimum. The particles get killed and respawned around the origin before they can reach the minimum at $(4,4)$. Meanwhile, \texttt{proximity~spawn} can \enquote{learn} from its current particle positions and therefore overcome a bad starting condition.}
    \label{fig:drought}
\end{figure}

It is here where we notice a difference between \texttt{static} and \texttt{proximity spawn}: When, we consider a paraboloid potential with minimum at $(4,4)$, i.\,e.\ at some distance from the initial spawn point $(0,0)$, \texttt{static spawn} does not reproduce the target distribution well: After being spawned around $(0,0)$, chains do not survive long enough to reach the minimum and provide good samples. For \texttt{proximity spawn}, however, this problem does not occur since new particles will spawn close to existing ones and the entire \enquote{swarm} thus may move to the minimum. We can see this effect in action by looking at kernel density estimates for said situation in Fig.~\ref{fig:drought}.

\begin{figure}
    \centering
    \includegraphics[scale=1]{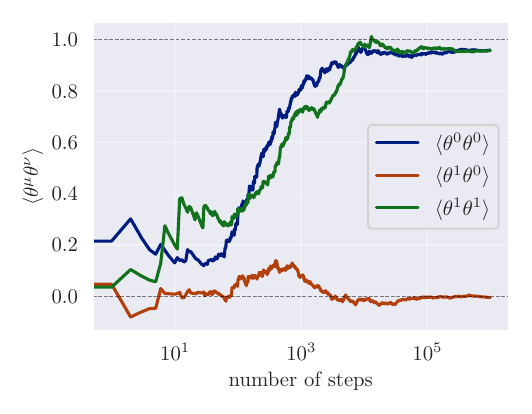}
    \caption{\textbf{$\textbf{2}$-dimensional Gaussian} --- Second moments of the ensemble as a function of iterations (approximately).
    As described in Section~\ref{subsec:equipartition}, the equipartition condition at unit (inverse) temperature: $\smash{\langle \theta^\mu \theta^\nu \rangle = \delta^{\mu \nu}}$ is approximately reached after about $1\%$ of the total number of iterations.
    \vspace{10pt}}
    \label{fig:equipartition}
\end{figure}

Finally, we test whether our sampler results adhere to the notion of equipartition in $n=2$ dimensions. More specifically, we want to check whether the (cumulative) moments follow \eqref{eq:equipartition}. The moments of all chains are displayed in Fig.~\ref{fig:equipartition}. Please note: While they are ordered by iteration steps of the algorithm, the $x$ axis is only roughly proportional to Monte Carlo time since the number of chains varies. Still, the graph resembles the earlier results \citep{10.1093/mnras/Rover} well: After approximately $1 \%$ of the total number of iterations, we reach the desired $\smash{\langle \theta^\mu \theta^\nu \rangle = \delta^{\mu \nu}}$.

\begin{figure*}[h!]
    \begin{minipage}{\linewidth}
    \centering
    \includegraphics*[scale=0.99]{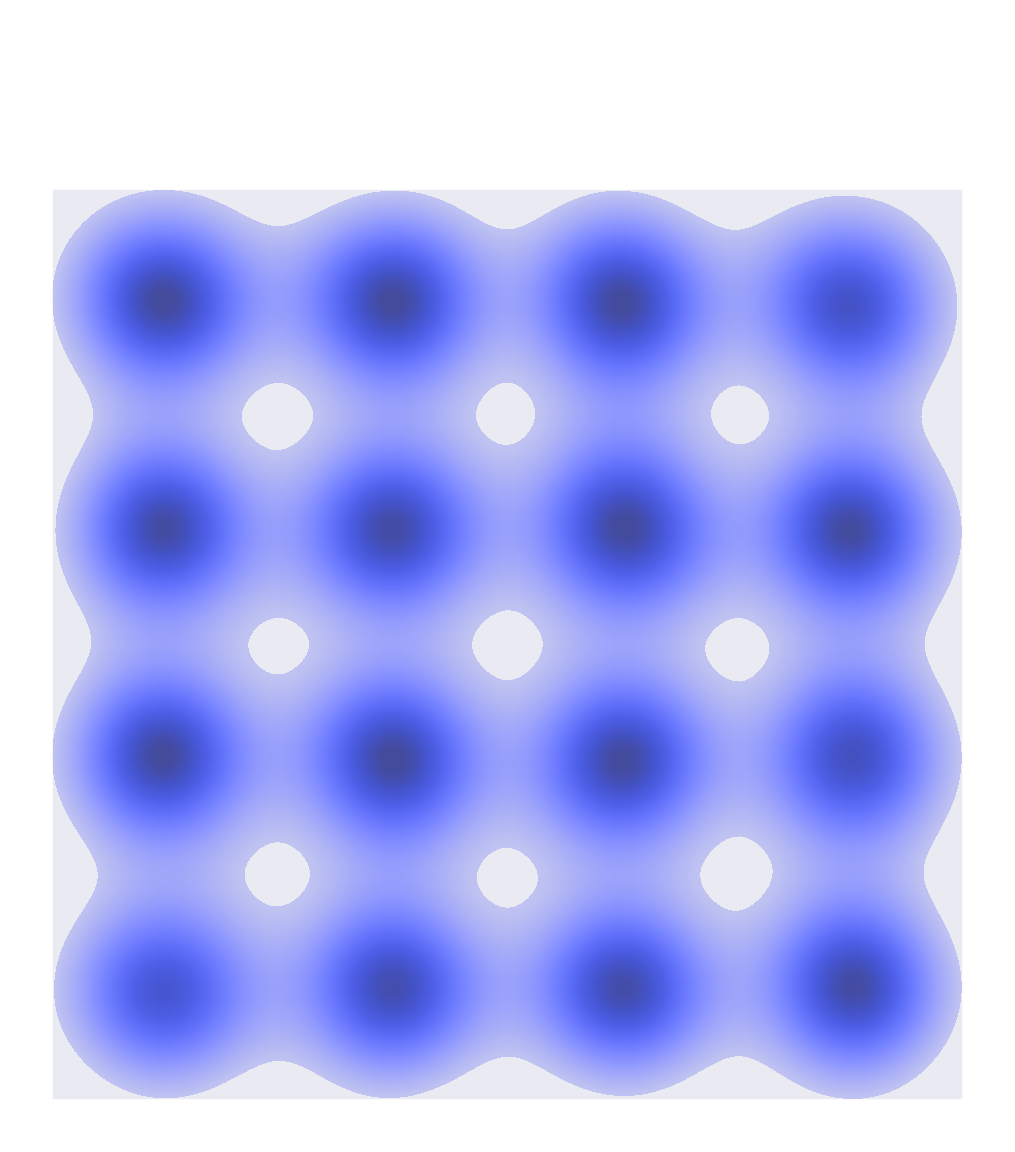}
    \includegraphics*[scale=0.99]{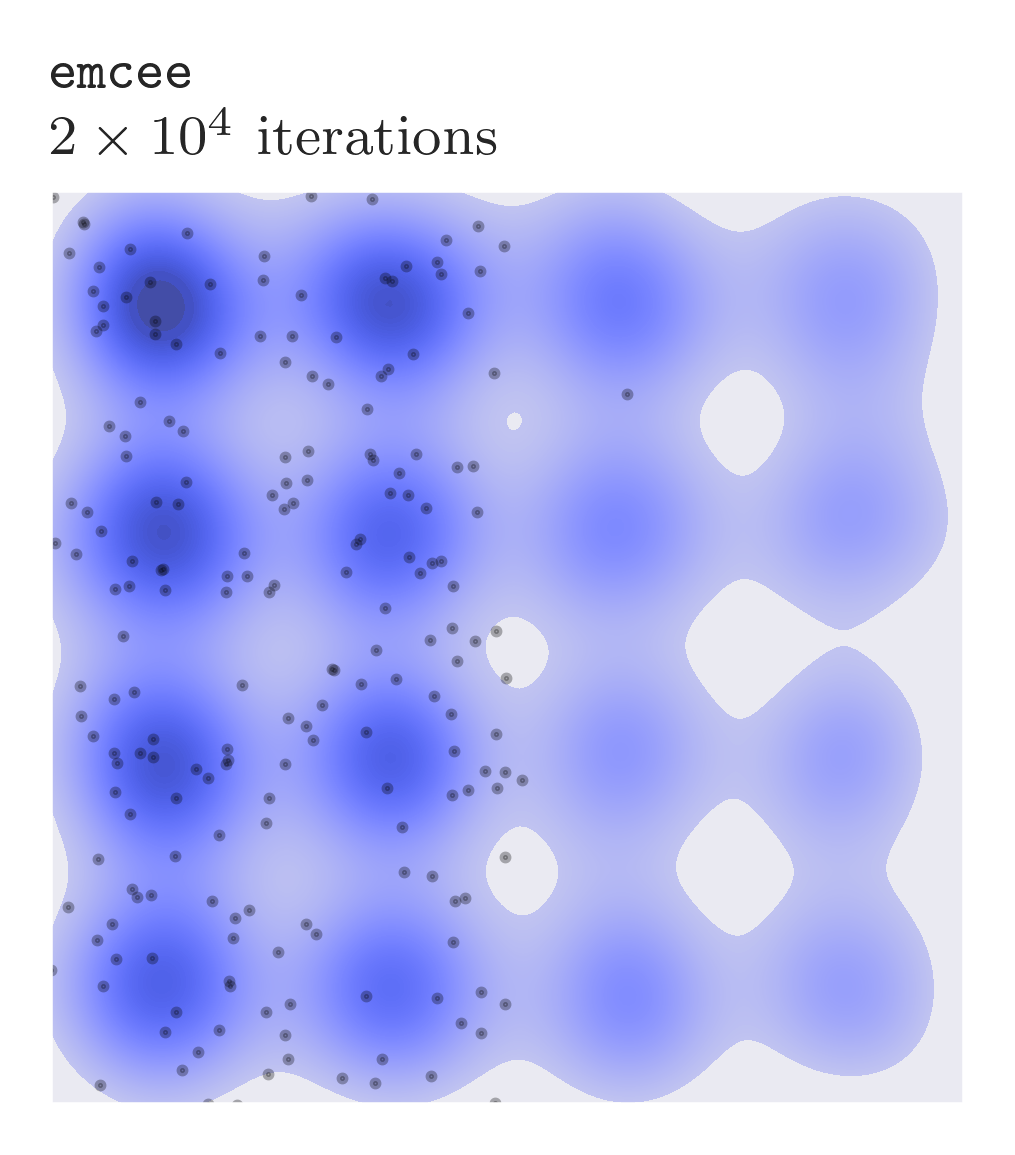}
    \includegraphics*[scale=0.99]{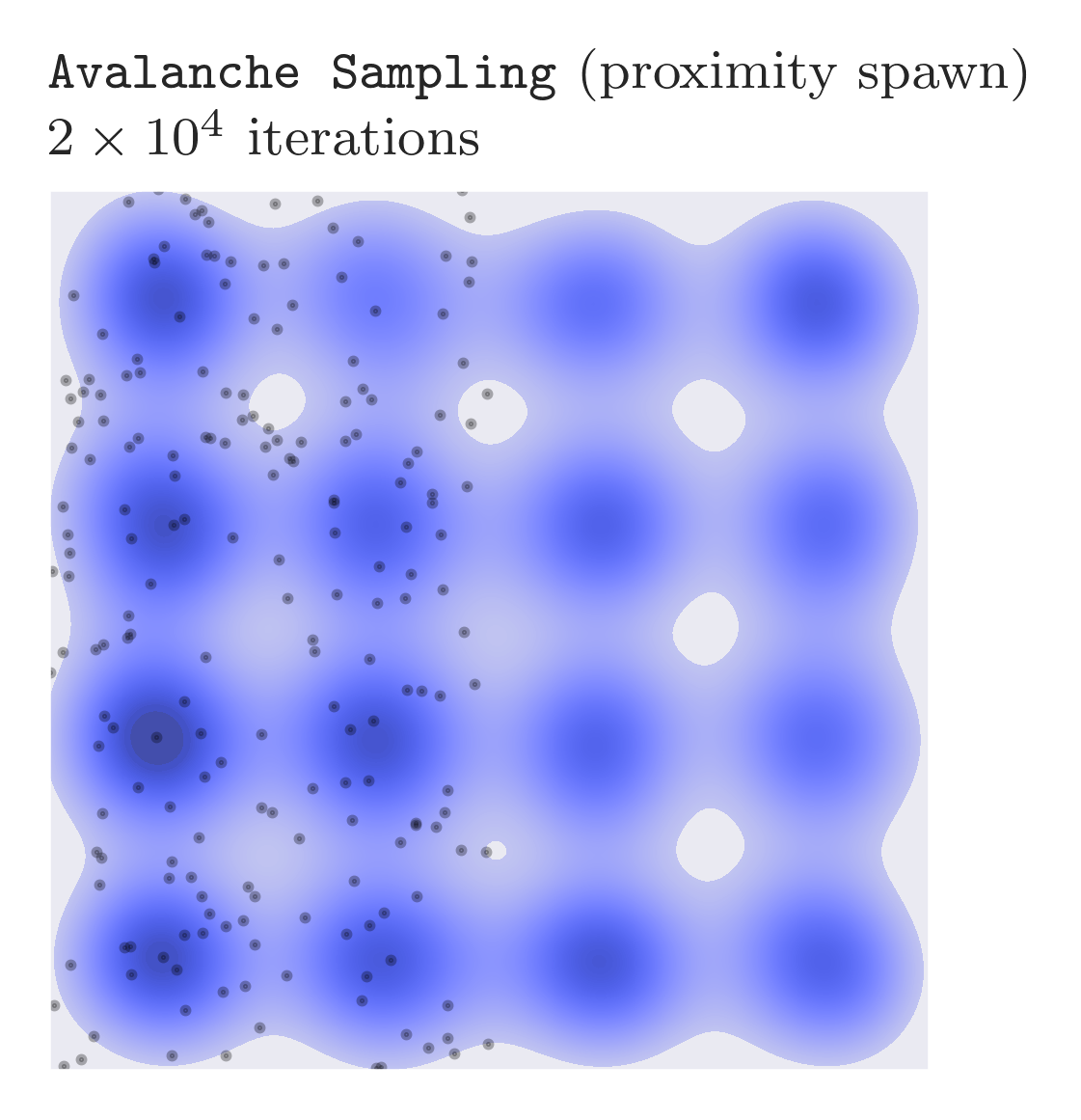}
    \caption{\textbf{\textit{Egg-carton}} --- Performance of \texttt{Avalanche Sampling} on the potential described in \eqref{egg-carton}. To make things worse, we initialise the chains for both methods only in the left half. As a reference, the leftmost plot is a reference sampling procedure using a long run with \texttt{emcee}.}
    \label{fig:test_vs_emcee_egg_carton}
    \end{minipage}
\end{figure*}

\begin{figure*}[h!]
    \begin{minipage}{\linewidth}
    \centering
    \includegraphics*[scale=0.99]{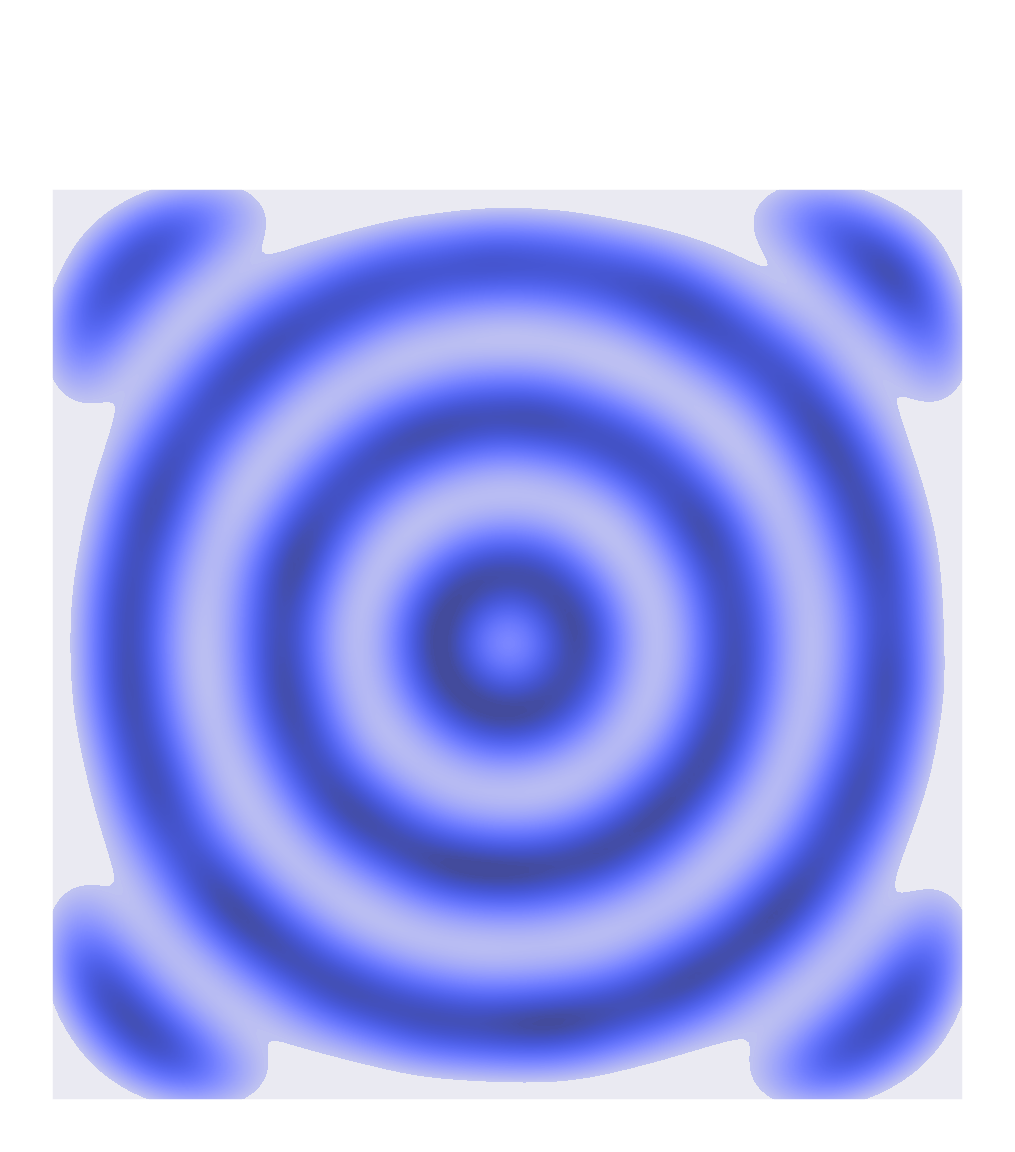}
    \includegraphics*[scale=0.99]{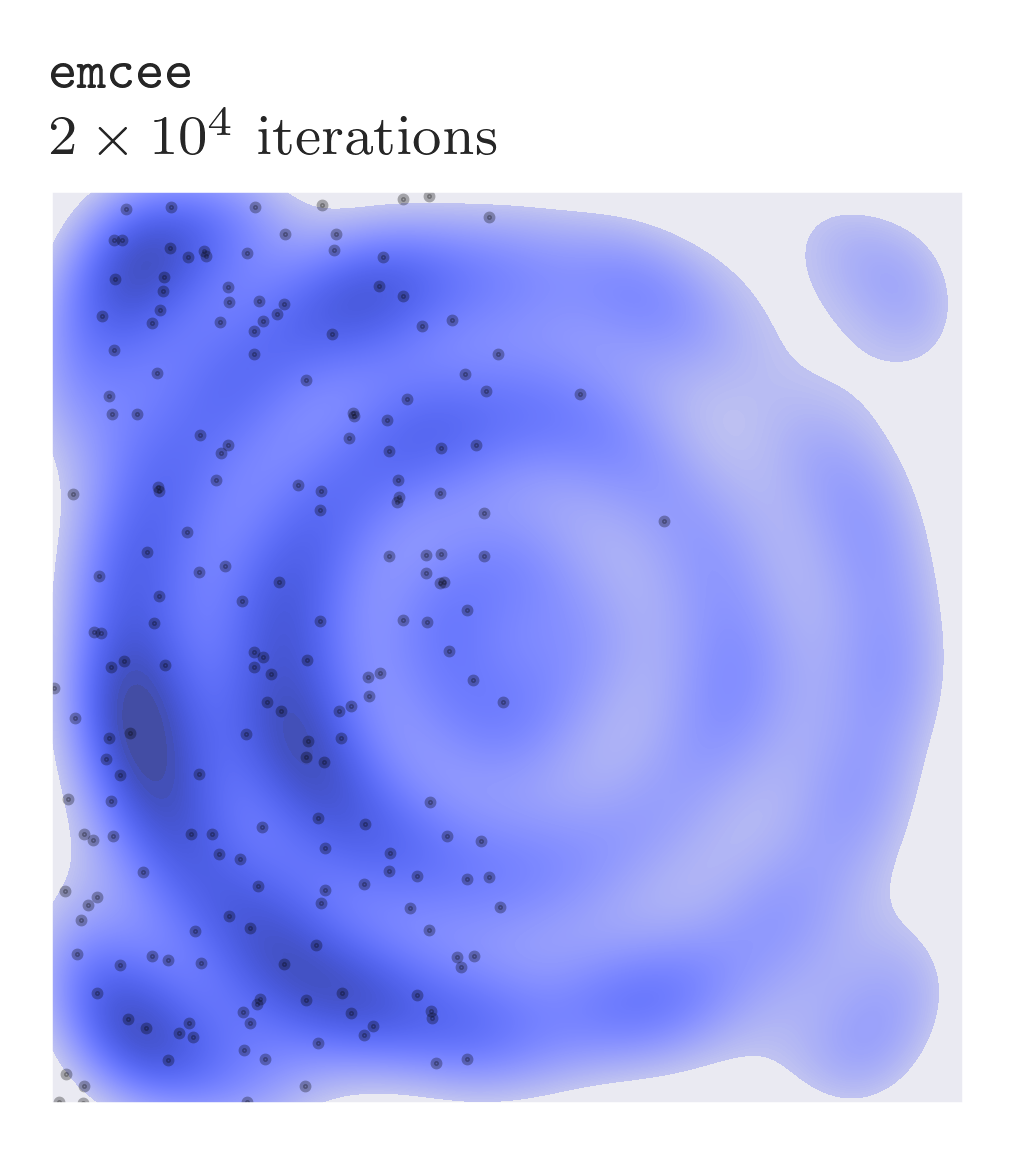}
    \includegraphics*[scale=0.99]{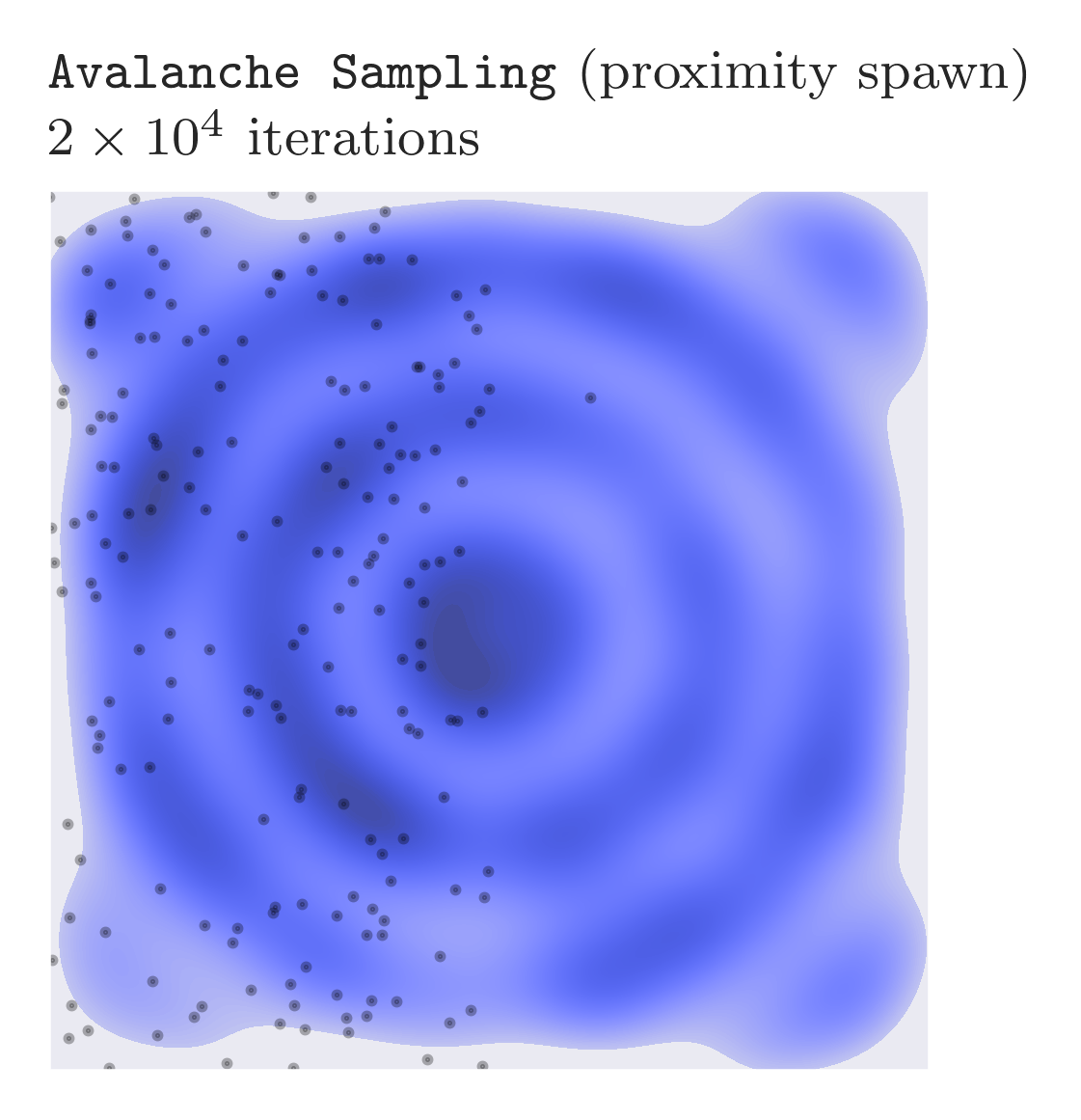}
    \caption{\textbf{\textit{Ripples} potential} --- Performance of \texttt{Avalanche Sampling} on the potential described in \eqref{eq:ripples}. To make things worse, we initialise the chains for both methods only in the left half. As a reference, the leftmost plot is a reference sampling procedure using a long run with \texttt{emcee}.}
    \label{fig:test_vs_emcee_ripple}
    \end{minipage}
\end{figure*}

\subsection{MULTIMODAL DISTRIBUTIONS}

\paragraph*{Setup.}
In order to test our ideas on a multimodal distribution, we now examine an \textit{egg-carton} potential given by
\begin{equation}
    \label{egg-carton}
    \frac{1}{2}\chi^2 (y \mid \theta) = \sum_{i=1}^n \sin^2(2 \pi \theta^i),
\end{equation}
as well as a \textit{ripples} potential given by
\begin{equation}
    \label{eq:ripples}
    \frac{1}{2} \chi^2 (y \mid \theta) = \sin(3 \pi |\theta|),
\end{equation}
both with a flat prior over $[-1,1]^n$
\begin{equation}
    \pi (\theta) = \begin{cases}
        2^{-n} & \text{if } \theta \in [-1,1]^n \\
        0 & \text{else}
    \end{cases}.
\end{equation}

\paragraph*{Results.}
In Figs.~\ref{fig:test_vs_emcee_egg_carton} and~\ref{fig:test_vs_emcee_ripple}, we compare the performance of \texttt{Avalanche Sampling} to a run with \texttt{emcee} \citep{emcee} on the egg-carton and ripples potential after a short run. However, we start in a non-ideal situation: We purposely initialise the chains only in the left half of the potential. The resulting kernel density estimates indicate, that in contrast to \enquote{fixed-$N$}-sampling methods, a macrocanonical approach to sampling can overcome an unfortunate starting condition, even if the proposal distribution is chosen unsuitably small. We believe this points to its robustness towards high-dimensional multimodal problems: \texttt{Avalanche Sampling} seems to make better use of the available space in a short amount of time, as the final samples are distributed more evenly across the potential.

In both cases, we were able to recover the evidence $p(y)$ estimated by a state-of-the-art dynamic nested sampling library, \texttt{dynesty} \citep{dynesty_2020}.

\subsection{COSMOLOGY: Application to supernova data}
\label{sec:supernovae}
\paragraph*{Setup.} As a topical  example for non-Gaussian likelihoods we reinvestigate constraints on the matter density $\Omega_m$ and the dark energy equation of state $w$ from the distance-redshift relation of supernovae of type Ia \citep{Riess1998, goobar_supernova_2011}. For simplification, we determine the luminosity distance for spatially flat FLRW-cosmologies with a dark energy equation of state parameter that is constant in time. Specifically, we derive constraints from the Union2.1-data set \citep{suzuki_hubble_2012, amanullah_spectra_2010, kowalski_improved_2008}: The distance modulus $y(a)$ as a function of the scale factor $a$ is given by
\begin{align} \label{eqn:DistModulus}
&y(a|\Omega_m,w) =  10 +  5 \log\left(\frac{1}{a}\, \int_{1}^{a}\dd{a^\prime}\:
\frac{1}{{a^\prime}^2 H(a^\prime)}\right)\; \text{with} \nonumber\\
& H(a^\prime) = H_0 \sqrt{\Omega_m {a^\prime}^{-3} + (1-\Omega_m) {a^\prime}^{-3(1+w)}}.
\end{align}
Constructing the likelihood for the two parameters $\Omega_m$ and $w$ for Gaussian errors $\sigma_i$ in the distance moduli $y_i$, we neglect correlations between the data points:
\begin{equation}
\likeli(y\mid\Omega_m,w) \propto \exp\left(-\frac{1}{2}\chi^2(y\mid\Omega_m,w)\right)
\end{equation}
with
\begin{equation}
\chi^2(y\mid\Omega_m,w) = \sum_i\left(\frac{y_i - y(a_i\mid\Omega_m,w)}{\sigma_i}\right)^2.
\label{eqn:MCMC_likelihood}
\end{equation}
We choose a uniform prior $\pi(\Omega_m,w)$ and subsequently probe the corresponding posterior distribution $p(\Omega_m, w \mid y)$ in the two-dimensional parameter space composed of $\Omega_m$ and $w$.

\begin{figure*}
	\centering
	\includegraphics[width =0.45 \textwidth]{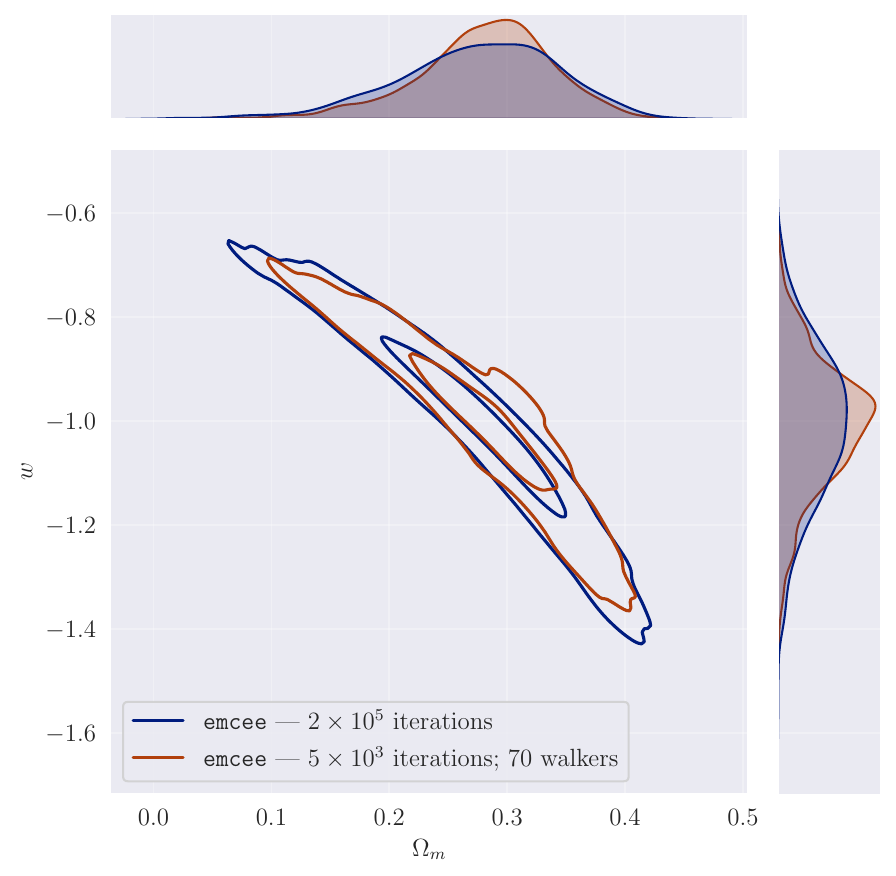}
	\includegraphics[width =0.45 \textwidth]{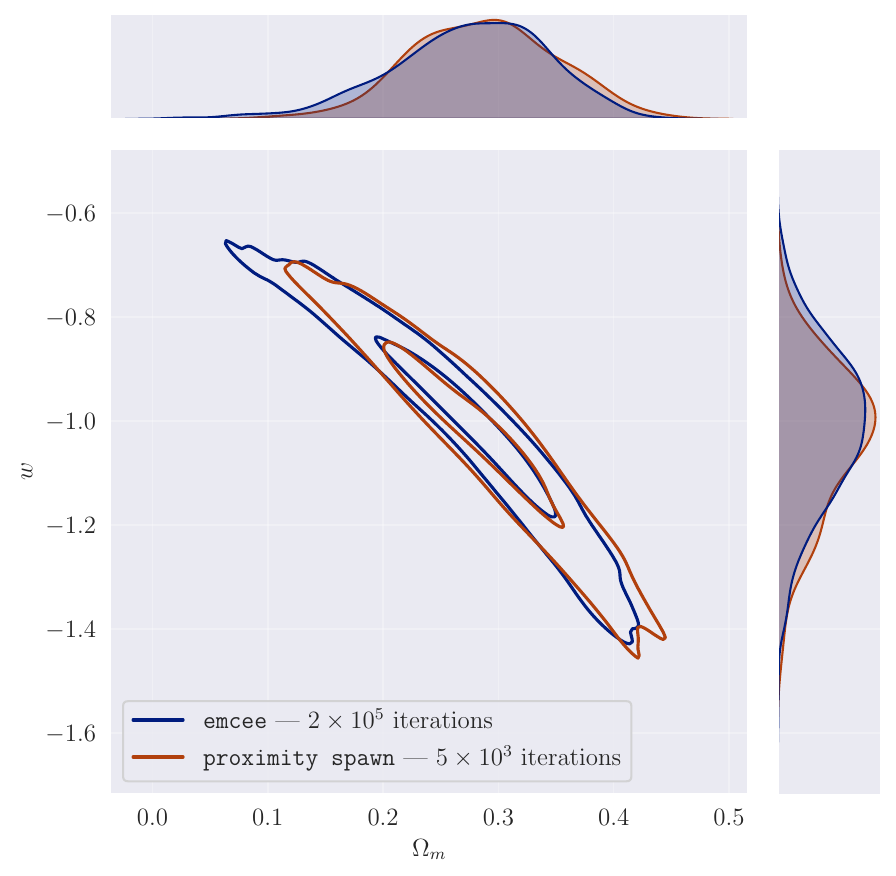}\\
    \caption{\textbf{Supernova likelihood} --- Test comparison of the \texttt{emcee} \citep{emcee} sampling method \textit{(left)} and \texttt{Avalanche Sampling} (\texttt{proximity spawn}) \textit{(right)}, both superimposed onto a reference long run using \texttt{emcee}. The chemical potential $\mu$ has been chosen such that the number of chains in equilibrium is comparable to the number of walkers in the short \texttt{emcee} run.}
    \label{comparison of methods}
\end{figure*}

\paragraph*{Results.} We compare the sampler distribution with a state-of-the-art Markov Chain Monte Carlo Ensemble sampler, called \texttt{emcee}, that utilises an affine-invariant sampling method \citep{emcee}. Fig.~\ref{comparison of methods} illustrates the sampling results of the \texttt{emcee} method and our \texttt{Avalanche Sampling} algorithm for an equal number of iterations. It is evident that the overall bend of the equipotential lines is reproduced more authentically by the macrocanical approach to sampling.
Again, we recover the evidence $p(y)$ from the number of chains $N$ as described in Section \ref{sec:distOfN}, with a value of $0.0306 \pm 0.0005$, while a Dynamic Nested Sampling algorithm, \texttt{dynesty} \citep{dynesty_2020}, yields $0.034 \pm 0.005$.

\section{summary and discussion}\label{sect_summary}
Subject of this paper was the extension of ensemble Markov-chains methods for statistical inference to the case where the number of chains is not fixed, but controlled by means of a chemical potential $\mu$, or equivalently, a fugacity $z = \exp(\beta\mu)$. From the point of view of statistical mechanics, this is equivalent to replacing the canonical by a macrocanonical ensemble, with a suitable partition function depending on inverse temperature $\beta$, the sources $\smash{J_\alpha}$ corresponding to the coordinates $\smash{\theta^\alpha}$, and the chemical potential $\mu$.

\begin{enumerate}
\item{The macrocanonical partition sum $\smash{\Xi[\beta,J_\alpha,\mu]}$ falls back on the Bayes-evidence $p(y)$ for unit (inverse) temperature $\beta = 1$, unit fugacity $z = \exp(\mu)$ and vanishing $J_\alpha$. Its logarithmic derivatives with respect to $\smash{J_\alpha}$ can be used to determine the cumulants of the posterior distribution $p(\theta|y)$, as in the the case of canonical partitions. In complete analogy, one would obtain results like the Cram{\'e}r-Rao inequality and the realisation of the minimum variance bound by the inverse Fisher-matrix for Gaussian distributions.}
\item{Differentiation of the Helmholtz free energy $\smash{F(\beta,J_\alpha) = -\ln Z[\beta,J_\alpha]/\beta}$ with respect to temperature yields the Shannon-expression for the information entropy of the posterior distribution, which is likewise the case for the macrocanonical potential $\smash{\Omega(\beta,J_\alpha,\mu)}$, up to a number of additional terms, originating from the Gibbs-factor $1/N!$. In addition, we find the macrocanonical information entropy to scale with the total number of Markov-chains in the ensemble in thermodynamic equilibrium.
Second logarithmic derivatives of $\Omega(\beta,J_\alpha,\mu)$ lead naturally to quantities analogous to the specific heat and the variance of $\smash{\chi^2/2+\phi}$, i.e. the analogue of the potential, over the allowed parameter space.
Statements about virialisation and equipartition are applicable to macrocanonical ensembles and canonical ensembles alike.}
\item{Differentiations of the macrocanonical potential $\Omega(\beta,J_\alpha,\mu)$ with respect to chemical potential $\mu$ result in the number of Markov-chains to be expected in equilibrium, but more importantly, provide a pathway to access the Bayes-evidence. In fact, $p(y)$ corresponds to $\bra N\ket$ in thermal equilibrium, and the numerical values obtained in this way correspond well to the results from other samplers. More importantly, perhaps, is the clarity concerning the fluctuation statistics of $N$, which we demonstrate in fact to follow a Poisson-distribution, where we have checked the typical relation $\bra N\ket = \bra N^2\ket^c$. This yields in turn a clearly defined error on the estimated Bayes-evidence.
However, we wish to point out that our method is not competitive with approaches designed solely for evidence determination, such as Nested sampling, in terms of speed and necessary compute. Indeed, \texttt{Avalanche sampling} is unique in that it generates an estimate for the evidence as a valuable byproduct, all while accomplishing its main task, sampling from the posterior probability distribution.}
\item{Varying the chemical potential $\mu$ provides a mechanism to control the sampling process, similarly to changing the temperature $\beta$ in simulated annealing, in the same way as high temperatures make it more likely to allow Markov-chains to propagate into disfavoured regions of parameter space, and as lowering the temperature accumulates Markov-chains in favourable regions. The choice of high chemical potentials would enable an exploratory mode while a low value would generate many Markov-chains at favourable regions of parameter space, making the sampling more efficient. At the same time, stationarity of the number of Markov-chains is an indicator of thermalisation and convergence.
However, finding optimal choices for $\mu$ is difficult, especially for high-dimensional parameter spaces. While $\mu$ does not change the acceptance rates, it exponentially changes the number of chains that need to be simulated. We therefore propose two different ways of choosing $\mu$ in a general setting. The first is to use a variant of Nested Sampling to estimate the Bayes-evidence and then choose $\mu$ such that the distribution of the number of chains in equilibrium is as desired. The second is to try and steer $\mu$ such that the amount of chains increases or decreases to the desired distribution. When it finally settles in, we could again extract the evidence, or decrease $\mu$ in a controlled manner to perform the before mentioned simulated annihilation procedure.}
\item{We provide the algorithm \texttt{Avalanche Sampling} as an implementation of macrocanonical sampling, verify its validity with test cases and ensure that its numerical results conform to our analytical calculations with $\smash{\Xi[\beta,J_\alpha,\mu]}$, in particular we demonstrate that the fluctuation statistics of the number of Markov-chains $N$ is correctly reproduced. Bayes-evidences computed from the expectation value of $N$ conform to the results from other algorithms.}
\item{\texttt{Avalanche Sampling} is able to explore parameter spaces quickly, even for fixed chemical potential which we investigated against other ensemble sampling algorithms in pathological examples: Specifically, we use test cases with many minima on a regular lattice and  potentials with concentric rings. We would argue that \texttt{Avalanche Sampling}'s proximity spawning, where a new Markov-chain is initialised close to an already existing chain, enables information exchange between the chains and reduces the necessary burn-in phase which is present in uniformly distributed spawning locations.}
\item{An actual application of macrocanonical sampling to the joint likelihood for the dark matter density $\Omega_m$ and the dark energy equation of state $w$ as it would be constrained by observations of supernovae of type Ia concludes our paper: We show that spawn and kill samples properly from the posterior distribution $\smash{p(\Omega_m,w|y)}$ and reaches fair sampling more quickly than other ensemble sampling methods without specific optimisation in either case. We would suspect, however, that macrocanonical sampling with a varying number of Markov-chains is to some extent degenerate with reweighting schemes.}
\end{enumerate}

In summary, we would like to emphasise that ensemble sampling methods with varying numbers of chains are well described by macrocanonical partition functions, which provide many insights into their information theoretical aspects. We intend to investigate how the efficiency of the sampling process depends on the choice of variables, $\beta,J_\alpha,N$ in the canonical case and $\beta,J_\alpha,\mu$ in the macrocanonical case, and identify pathways to further increase sampling efficiency.

\section*{Acknowledgements}
We would like to thank U. K{\"o}the for the suggestion of calling the macrocanonical sampling method \enquote{\texttt{Avalanche Sampling}}.

\paragraph{Funding information}
This work was supported by the Deutsche Forschungsgemeinschaft (DFG, German Research Foundation) under
Germany's Excellence Strategy EXC 2181/1 - 390900948 (the Heidelberg STRUCTURES Excellence Cluster).

\paragraph{Data availability}
Our python-implementation of the macrocanonical sampler \texttt{Avalanche Sampling}, \texttt{avalanchepy} is available on GitHub:
\url{https://github.com/cosmostatistics/avalanchepy}

\appendix
\section*{Acceptance probabilities}\label{acceptance probabilities}

When describing microscopic events, time reversibility is often considered a given.
Examples such as the Newton and the Schrödinger equation omit a $T$-symmetry in the
absence of external magnetic fields and in an inertial frame of reference. This
means that if $x(t)$ is a solution, then $x(-t)$ is one as well.
Accordingly, one would anticipate the presence of macroscopic consequences, such as the principle of detailed balance: In equilibrium
events are balanced by their reverse events.

Now, we turn back to the formulation of \cite{metropolis_equation_1953} and \cite{Hastings_1970} and view our algorithm as a Markov process, i.\,e.\@ a series of states 
\begin{equation}
\theta \rightarrow \theta' \rightarrow \theta'' \rightarrow \theta''' \rightarrow \ldots
\end{equation}
of some system where the current state only depends on the preceding state.
Any such process is uniquely defined by
the system's to transition probability $p(\theta' \mid \theta)$ from $\theta$ to $\theta'$. In the
limit of infinitely many iterations, the states will eventually be distributed
according to some stationary probability distribution, which is chosen to take a certain form, that of the \emph{target distribution}
$p^*(\theta)$. If $\theta$ is already distributed according to $p^*$ at a certain time, then one finds the eigenvector equation
\begin{equation}
\sum_\theta p(\theta' \mid \theta) p^*(\theta) = p^*(\theta')
\end{equation}
as a requirement for the transition probability. Aside from the normalisation condition $\sum_\theta p(\theta \mid \theta') = 1$, the transition probability should be ergodic and irreducible, i.\,e.\@ it should be possible to reach any state from any other state in a finite number of steps. This ensures that the system will eventually reach equilibrium.
There are multiple ways to construct a transition probability $p$ that fulfils
the above eigenvector equation. We choose the condition of \emph{detailed balance}
\begin{equation}
    \label{eq:detailed_balance}
    p(\theta' \mid \theta) p^*(\theta) = p(\theta \mid \theta') p^*(\theta').
\end{equation}
One may easily see that this indeed implies the eigenvector equation.

In concrete algorithms, moving from a state $\theta$ to $\theta'$ is done in two
steps: First, a \emph{proposal} for a new state is drawn from some proposal
distribution, $\theta' \sim T(\theta' | \theta)$. Secondly, the proposal
is \emph{accepted} with a probability given by the acceptance distribution
$\alpha(\theta', \theta)$.
The total transition probability may thus be
written as the product of the two,
$p(\theta'| \theta) = T(\theta'|\theta\:) \alpha(\theta', \theta)$.
With this, we may write the detailed balance condition
\eqref{eq:detailed_balance} as
\begin{equation}
    \frac{\alpha(\theta', \theta)}{\alpha(\theta, \theta')} = \frac{T(\theta\; \mid \theta')p^*(\theta')}{T(\theta' \mid \theta\;)p^*(\theta\;)}.
\end{equation}
As we want the acceptance probability to be bounded by $1$, a sensible choice
for $\alpha$ would be
\begin{equation} \label{eq:alpha_2}
    \alpha(\theta', \theta) = \min\left\{1, \frac{T(\theta\; \mid \theta')p^*(\theta')}{T(\theta' \mid \theta\;)p^*(\theta\;)} \right\}.
\end{equation}

For the case of the concrete \texttt{Avalanche} algorithm, the acceptance probabilites were calculated from the proposal probabilities \eqref{eq:proposeSpawnStatic}-\eqref{eq:proposeKillProximity}, using Hastings formula \eqref{eq:alpha}. The detailed expressions are given below.

For \texttt{static spawn}, we get
\begin{align*}
    \textbf{spawn:}&&\;\alpha(^* \theta_{N+1}) &= \min \left\{ 1, \exp\left[\mu - \frac{1}{2}\chi^2(y|\theta_{N + 1}) - \phi(\theta_{N + 1})\right] \cdot \left[(N+1)\; p(\theta_{N+1})\right]^{-1} \right\}\\
     \textbf{kill:}&&\;\alpha(^\dagger \theta_k) &= \min \left\{ 1, \exp\left[-\mu + \frac{1}{2}\chi^2(y|\theta_{k}) + \phi(\theta_{k})\right] \cdot N\; p(\theta_{k}) \right\}.
\end{align*}

For \texttt{proximity spawn}, we get
\begin{align*}
    \textbf{spawn:}&&\;\alpha(^* \theta_{N+1}) &= \min\left\{  1, \exp\left[\mu - \frac{1}{2}\chi^2(y|\theta_{N + 1}) - \phi(\theta_{N + 1})\right] \cdot N
    \;\left[\sum_{i=1}^N \sum_{\substack{j=1, j\neq i}}^N p(\theta_i \mid \theta_j)\right]^{-1}
    \right\}\\
     \textbf{kill:}&&\;\alpha(^\dagger \theta_k) &= \min\left\{1, \exp\left[-\mu + \frac{1}{2}\chi^2(y|\theta_{k}) + \phi(\theta_{k})\right]  \cdot \frac{1}{N-1} \; \sum_{i=1}^N \sum_{\substack{j=1,j\neq i}}^N p(\theta_i \mid \theta_j)
     \right\}.
\end{align*}
It is noteworthy that the sum in \eqref{eq:proposeKillProximity} drops out, making the computation easier by a large margin. \vspace{10pt}

\bibliography{references}

\end{document}